\PassOptionsToPackage{unicode}{hyperref}
\PassOptionsToPackage{hyphens}{url}
\documentclass[
]{article}
\usepackage{xcolor}
\usepackage{todonotes}
\usepackage{braket}
\usepackage{amsmath,amssymb}
\usepackage{iftex}
\ifPDFTeX
  \usepackage[T1]{fontenc}
  \usepackage[utf8]{inputenc}
  \usepackage{textcomp} 
\else 
  \usepackage{unicode-math} 
  \defaultfontfeatures{Scale=MatchLowercase}
  \defaultfontfeatures[\rmfamily]{Ligatures=TeX,Scale=1}
\fi
\usepackage{lmodern}
\ifPDFTeX\else
\fi
\IfFileExists{upquote.sty}{\usepackage{upquote}}{}
\IfFileExists{microtype.sty}{
  \usepackage[]{microtype}
  \UseMicrotypeSet[protrusion]{basicmath} 
}{}
\makeatletter
\@ifundefined{KOMAClassName}{
  \IfFileExists{parskip.sty}{%
    \usepackage{parskip}
  }{
    \setlength{\parindent}{0pt}
    \setlength{\parskip}{6pt plus 2pt minus 1pt}}
}{
  \KOMAoptions{parskip=half}}
\makeatother
\usepackage{longtable,booktabs,array}
\usepackage{calc} 
\usepackage{etoolbox}
\makeatletter
\patchcmd\longtable{\par}{\if@noskipsec\mbox{}\fi\par}{}{}
\makeatother
\IfFileExists{footnotehyper.sty}{\usepackage{footnotehyper}}{\usepackage{footnote}}
\makesavenoteenv{longtable}
\usepackage{graphicx}
\makeatletter
\newsavebox\pandoc@box
\newcommand*\pandocbounded[1]{
  \sbox\pandoc@box{#1}%
  \Gscale@div\@tempa{\textheight}{\dimexpr\ht\pandoc@box+\dp\pandoc@box\relax}%
  \Gscale@div\@tempb{\linewidth}{\wd\pandoc@box}%
  \ifdim\@tempb\p@<\@tempa\p@\let\@tempa\@tempb\fi
  \ifdim\@tempa\p@<\p@\scalebox{\@tempa}{\usebox\pandoc@box}%
  \else\usebox{\pandoc@box}%
  \fi%
}
\def\fps@figure{htbp}
\makeatother
\ifLuaTeX
  \usepackage{luacolor}
  \usepackage[soul]{lua-ul}
\else
  \usepackage{soul}
\fi
\ifPDFTeX
  \newcommand{\RL}[1]{\beginR #1\endR}
  \newcommand{\LR}[1]{\beginL #1\endL}
  \newenvironment{RTL}{\beginR}{\endR}
  \newenvironment{LTR}{\beginL}{\endL}
\fi
\ifluatex
  \newcommand{\RL}[1]{\bgroup\textdir TRT#1\egroup}
  \newcommand{\LR}[1]{\bgroup\textdir TLT#1\egroup}

\fi
\usepackage{bookmark}
\IfFileExists{xurl.sty}{\usepackage{xurl}}{} 
\makeatletter
\@ifundefined{xmpquote}{}{}
\makeatother
\hypersetup{
  pdftitle={Designing a Hybrid Digital/Analog Quantum Physics Emulator as Open Hardware},
  hidelinks,
  pdfcreator={LaTeX via pandoc}}

\title{\protect\phantomsection\label{h.x9ax18y6g4n7}{}Designing a Hybrid Digital/Analog Quantum Physics Emulator as Open Hardware}
\usepackage{etoolbox}
\makeatletter
\providecommand{\subtitle}[1]{
  \apptocmd{\@title}{\par {\large #1 \par}}{}{}
}
\makeatother
\author{Marcus Edwards \\ Member, IEEE}

\begin{document}
\maketitle

\subsection{Abstract}\label{abstract}

Existing approaches to emulating quantum computing algorithms using
classical electronic hardware are limited by exponential scaling
limitations in space, such as circuit size, or time, such as runtime or
bandwidth. We introduce a scheme for representing quantum information
using analog signals that lessens the bandwidth limitation problem (in certain regimes) seen
in existing approaches \cite{cour_signal_based_2015, la_cour_classical_2016} by taking full advantage of the
ability of analog signals to encode information using RMS voltage as well as frequency and phase. 
We introduce the mathematical framework for
this representation, which separates the information relevant for
measurement in the computational basis from information that is not
relevant to it. We introduce circuits that take advantage of this
separation of concerns to achieve simplifications, for working with
quantum information in this representation. We argue that it is
comparatively very inexpensive (as low as \textasciitilde\$5.00 / qubit)
to outmatch the computing capabilities of existing FPGA based emulators
\cite{pilch_fpga_based_2019}, though scaling beyond tens of qubits is still impractical due
to constraints of analog hardware module precision. However, our
approach opens the door to a new avenue by which classical emulators can
hope to improve: by improving on analog electronic circuit performance.

\subsection{Introduction}\label{introduction}

Meaningful quantum algorithms are difficult or impossible to simulate
using classical computers. This is both why we are interested in
building quantum computers that can perform these meaningful tasks, and
a challenge for quantum computing researchers that want to understand
how they work.

The first analog signal-based emulation of a universal quantum computer
was performed by La Cour et. al. \cite{cour_signal_based_2015}. Since then, more analog
emulators \cite{la_cour_classical_2016} and solvers \cite{pomorski_towards_2024} of quantum physics such as
Ginzburg--Landau and Schrödinger equations have been proposed. But do
these approaches overcome the exponential scaling problems that
classical digital simulations face, and the curse of dimensionality?

The problem is still present in these approaches. Consider that in the
first scheme \cite{cour_signal_based_2015}, a signal duration of the approximate age of the universe (13.77 billion years)
could accommodate only about 95 qubits. For most meaningful quantum
algorithms, that isn\RL{'}t nearly enough. The electrical emulators that
have been developed simply don\RL{'}t scale with complexity of important
problems.

Digital hardware emulators have also been proposed, for example based on
FPGAs. These approaches do not claim to get around the problem of
exponential scaling, but instead show that it is possible to shift where
this problem has its impact. For example, the early approach of Pilch
and Dlugopolski \cite{pilch_fpga_based_2019} shifts the scaling to a problem of the cost and
nature of the FPGA technology used.

While quantum computing offerings from companies around the world remain
in the Noisy Intermediate Scale Quantum (NISQ) regime, it is important
that researchers can still somehow gain insights into how potential
quantum computing algorithms and applications may look. For this
purpose, many companies offer quantum simulators. Quantum emulators may
be a desirable enhancement to simulators which otherwise run on hardware
which is less specialized to the problem of quantum computer emulation
and therefore less efficient.

When considering optimal choices for technology at each layer of the
stack in this paper, we prioritize the optimization of our proposed
solution as a function of economic cost and engineering feasibility.

\subsubsection{The Naive Hardware Scaling
Problem}\label{the-naive-hardware-scaling-problem}

Let us consider a naive approach to mapping our identification numbers
for quantum states directly to a set of simple sinusoidal analog
signals.

Let \(K\) be a weighted sum of \(N\) pure, single qubit quantum states
denoted \(\psi_{n}\). \(K\) can be given in terms of the trace of a
product of square matrices \(M\) and \(W\) of rank \(N\).

\[K = \overset{N}{\sum_{n = 0}}c_{n}|\psi_{n}\rangle = Tr(MW)\]

This is analogous to a continuous interpretation of the sum of the
diagonal elements of the matrix

\[K\overset{\cdot}{=}\int_{0}^{N}w(n)|\psi_{n}\rangle dn\]

where \(w:\mathbb{Z} \rightarrow \mathbb{C}\) is a function approximated
by the eigenvalues of the matrix \(W\) such that \(K\) is then a sum of \(N\) oure quantum states.

\(W|n\rangle \cdot \left\lbrack \begin{array}{r}
1 \\
1 \\
 \vdots \\
1
\end{array} \right\rbrack\overset{\cdot}{=}w(n)\).

To demonstrate a scalability issue with directly mapping sums of pure
states like \(K\) to a discrete set of analog signal representations,
assume that each pure quantum state is represented by an analog signal,
a sinusoid \(sin(n \cdot d\omega \cdot x)\).

Let us consider \(M\) and \(W\) to define a minimal set of hardware
components needed to emulate the state being represented. The rank of
\(M\) is equal to the number of individual sinusoids that need to be
concurrently generated. Therefore \(rank(M)\) is the number of wave
sources, or oscillators, required. The components of \(W\) each
represent a weight in the sum of the sinusoidal basis states.
Therefore \(rank(W)\) is the number of amplifiers required. Finally, the
number of two-way summers required to create the final ensemble is equal
to the number of sinusoids being summed, minus one: \(rank(M) - 1\).

This yields an expression of the complexity \(C\) of a direct and purely
hardware-based implementation of a simplified state emulation that
behaves in the way that \(K\) has been described:

\[c = 2 \cdot rank(M) + rank(W) - 1 = 3N - 1\]

We can see that 5 hardware components would be required to emulate a
qubit state by asserting that \(N = 2\), to account for each of the basis
states \(|0\rangle\) and \(|1\rangle\). If we introduce multi-qubit
states then we might revise the formula for to account for the unique
combinations of each qubit\RL{'}s basis states:

\[C = 3 \cdot 2^{Q} - 1\]

where \(Q\) is the number of qubits. This expresses the fact that the
task of directly simulating this simplified qubit state scales in
hardware resources with \(O(2^{n})\).

\subsubsection{Bandwidth Limitation
Problem}\label{bandwidth-limitation-problem}

In an existing emulation approach \cite{cour_signal_based_2015}, La Cour and Ott described an
implementation scheme for a signal-based emulation of general quantum
computing. This model was demonstrated using analog electronics. Their
scheme introduced a mapping from quantum states to electrical analog
phase representations.

The model starts with representing the quantum state \(|0\rangle\) by
the in-phase and quadrature components of an analog electrical signal.
The \(|0\rangle\) state is defined as \(s(t)\):

\[s(t) = a \cdot cos(\omega_{c}t) - bsin(\omega_{c}t) = Re\lbrack\alpha e^{i\omega_{c}t}\rbrack\]

This state can be represented by a sinusoidal analog electronic signal
\(\alpha\). \(a\) then represents the real part of the sinusoidal signal
and \(b\) the imaginary part. The in-phase and quadrature amplitudes can
be obtained by multiplying the state by in-phase and quadrature
reference signals and applying a low-pass filter with a bandpass below
\(2\omega_{c}\).

This can also be extended to model a general single qubit state,
\(s(t) = Re\lbrack\psi(t)e^{i\omega_{c}t}\rbrack\). Let
\(\psi(t) = \psi_{R}(t) + i\psi_{I}(t) = \alpha_{0}e^{i\omega t} + \alpha_{1}e^{- i\omega t}\),
a combination of the basis states \(|0\rangle\) and \(|1\rangle\). Then
we can redefine \(s(t)\) to be a general one-qubit state:

\[s(t) = \psi_{R}(t) \cdot cos(\omega_{c}t) + \psi_{I}(t) \cdot sin(\omega_{c}t)\]

This achieves a way by which a general state
\(|\psi\rangle = \psi_{t} = \alpha\) can be modelled. The real and
imaginary parts of \(\psi\) serve as the in-phase and quadrature
components of the carrier signal. In-phase and quadrature references are
used in the following configuration, where \(\otimes\) represents a
4-quadrant multiplier. A 4-quadrant multiplier circuit produces the
product of its input voltages and either input voltage may be positive
or negative. The \(- \frac{\pi}{2}\) phase shift provides the quadrature
reference, and a positive analog filter is used to finally acquire the
state \(s(t)\).

\begin{center} 
\includegraphics[width=4.55157in,height=2.53273in,alt={s\_t.png}]{s\_t.png}
\end{center}

Then the analog components required to represent a quantum state are:

\begin{itemize}
  \item 2 analog sources
  \item 2 quadrature multiplier circuits
  \item 1 phase shift circuit
  \item 1 bandpass filter
\end{itemize}

Similarly, \(n\) qubits with \(n\) complex coefficients can be expressed
in general using the formula

\[\psi(t) = \overset{2^{n} - 1}{\sum_{x = 0}}a_{x}\phi_{x}(t)\]

Each basis state corresponds to one of \(2^{n}\) frequencies. Each of
\(2^{n}\) complex components may be multiplied by their associated basis
states, and the resulting products summed. This can be achieved using
similar circuitry to that given in the example for \(s(t)\). Inner
products may be computed via time averaging.

It is clear that this is not efficient, and La Cour and Ott do not stop
here, instead outlining some more efficient approaches for addressing
and manipulating qubits of this sort. However, this outline demonstrates
for us that there is a fundamental bandwidth limitation problem.

Since a quantum state of n qubits is represented using a complex
oscillating time-domain signal, the number of qubits that can be encoded
is limited by the attainable bandwidth. Another limitation to consider
is the requirements for physical components. The proposed device
consists of only three types of electrical components: 4 quadrant
multipliers, operational amplifiers and analog filters. A gate uses a
fixed number of multipliers, adders and inverters per qubit. La Cour and
Ott claim \cite{cour_signal_based_2015} that the total number of components needed for the
implementation of a gate scales quadratically with the number of qubits
the gate operates on. La Cour and Ott estimate \cite{cour_signal_based_2015} that in order to achieve
a density of electrical circuitry footprint that scales exponentially
with the number of qubits, transistor density would need to improve by a
factor of 1000 from what it was in 2015. For context, it improved by
closer to 10x between 2015 and 2020 \cite{moore_state_2022}. If this goal were reached,
and encoding information with a 1 THz bandwidth were possible, they
claim \cite{cour_signal_based_2015} it would then be feasible to emulate a system of 40 qubits, which
is comparable to a high performance computer with 1 TB of RAM.

Due to the bandwidth limitation, the inefficiencies of this
implementation largely exist in the time domain. The time dependence of
a state introduces a relationship between the signal duration \(T\) and
number of modellable qubits \(n\). A signal duration of 10 hours would
yield roughly 50 qubits, while 1 year would yield roughly 60 qubits.
Even if \(T\) were on the order of the age of the Universe only about 95
qubits could be represented \cite{cour_signal_based_2015}.

La Cour and Ott conclude \cite{cour_signal_based_2015} that a quantum emulation device with an octave
spacing of qubit frequencies would be constrained by an exponential
scaling of required bandwidth. So, this signal based emulation
methodology also scales with untenable complexity.

\subsection{Dual Oscillator Representation
Scheme}\label{dual-oscillator-representation-scheme}

To overcome these downsides, we will create a new scheme with the
following properties:

\begin{itemize}
  \item A pair of oscillators must be suffcient to emulate \(n\) superimposed states
  \item The time required to perform a measurement of a state must not scale poorly with the complexity of the state
  \item A fixed set of hardware components must be sufficient to emulate a system of a significant number of qubits
  \item At least as much must be measurable about an emulated quantum state as is expected to be measurable in a theoretical quantum computing system
\end{itemize}

In order to build to a viable solution, we will first define the
elements of the computational space we are interested in representing.

A probabilistic state with \(n\) observable values can be modelled by a
probability simplex, a tetrahedron in \(n - 1\) dimensions. For example,
in the case of a qubit, the probability of measuring a \(|0\rangle\)
versus a \(|1\rangle\) can be expressed as a point on a line between two
endpoints. The endpoints represent states which are 100\% likely to
yield an observable when measured. A line is then the geometric
embodiment of the spectrum of the probabilistic mixtures of a
qubit\RL{'}s observable values. This representation loses important
information about the qubit however, since a qubit is not simply a pbit.
Every qubit state \emph{can} be mapped to a corresponding point on a
pbit simplex, but we can not map a point on the pit simplex back to a
unique and fully determined qubit state \cite[p. 12]{edwards_introduction_2026}.

The Bloch sphere provides a more true representation of a qubit state \cite[p. 5]{Majidy_Wilson_Laflamme_2024}
since we know that a qubit does behave the same as a pbit. Rather, in
order to fully model a qubit, it is important to retain knowledge of the
square root of its probability in a way that is not arbitrary. Two
dimensions are added on top of the probability simplex in order to
account for the sign of the square root of the probability, and for
imaginary values.

\begin{center} 
\includegraphics[width=4.86111in,height=1.875in,alt={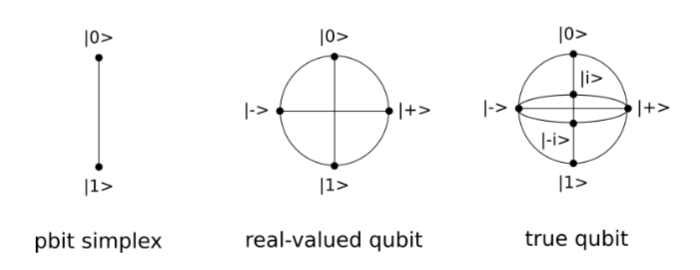}]{breakdown.png}
\end{center}

The imaginary component or sign of a qubit state does not affect the
probability of its observable values being measured. These do however
become necessary for the application of certain unitary operations like
\(Z\) and \(Y\) \cite[p. 61]{edwards_introduction_2026}.

Each coefficient in the equation for a quantum state can be broken into
three pieces of information:

1. Whether the coefficient is real or imaginary

2. Whether the coefficient is positive or negative

3. The magnitude of the coefficient

Items one and two are binary in nature, and can be captured by a bit
string of binary flags. The third item, however, exhibits a spectrum of
possibilities.

We shall distinguish between four sets of information processors. The
sum of these four subsystems will represent a full qubit state. The
first two subsystems will be responsible for maintaining information
directly relevant to a measurement. Their hardware should not be
required to interact with any other subsystems in order to answer a
measurement with an observable value. The third and fourth subsystems
will maintain information that is necessary to maintain about a quantum
state, but not relevant in the context of a measurement. Let the third
and fourth hardware modules together be named the ``flag processor
module'' and each maintain two pieces of information about a term in the
coefficients of each ket in a quantum state: whether it is imaginary and
whether it is negative.

In the case that the entire system is maintaining a single qubit of
quantum information, then the effects of Pauli operations on each
subsystem in the flag processor module can then be captured
by a simple state diagram.

\begin{center} 
\includegraphics[width=3.94884in,height=3.00801in,alt={state\_machine.png}]{state\_machine.png}
\end{center}

This state machine can be described by Boolean logic and clearly lends
itself to a trivial digital implementation.

The output state $c_{o_0}c_{o_1}c_{o_2}$ after a $-C$ transition from an input state $c_{i_0}c_{i_1}c_{i_2}$ is given by the following equations:

\begin{equation}
  c_{o_2} = \bar{c_{i_2}}
\end{equation}
\begin{equation}
  c_{o_1} = c_{i_0}
\end{equation}
\begin{equation}
  c_{o_0} = c_{i_1}
\end{equation}

In the case of a $Y$ transition, we have the following equations:

\begin{equation}
  Y_{o_2} = \bar{Y_{i_1}}(Y_{i_0}+Y_{i_2}) + Y_{i_0}Y_{i_1}\bar{Y_{i_2}}
\end{equation}
\begin{equation}
  Y_{o_1} = Y_{i_0}Y_{i_1}\bar{Y_{i_2}} + Y_{i_0}\bar{Y_{i_1}}\bar{Y_{i_2}}
\end{equation}
\begin{equation}
  Y_{o_0} = \bar{Y_{i_0}}Y_{i_2} + Y_{i_0}\bar{Y_{i_1}}\bar{Y_{i_2}} + Y_{i_1}Y_{i_0}Y_{i_2}
\end{equation}

In the case of $Z$:

\begin{equation}
  Z_{o_2} = Z_{i_0}Z_{i_1}Z_{i_2} + Z_{i_0}\bar{Z_{i_1}} + \bar{Z_{i_0}}\bar{Z_{i_1}}\bar{Z_{i_2}}
\end{equation}
\begin{equation}
  Z_{o_1} = Z_{i_1}\bar{Z_{i_2}} + \bar{Z_{i_0}}\bar{Z_{i_1}}\bar{Z_{i_2}} + Z_{i_0}\bar{Z_{i_1}}Z_{i_2}
\end{equation}
\begin{equation}
  Z_{o_0} = \bar{Z_{i_0}}(Z_{i_1} \oplus Z_{i_2}) + Z_{i_0}Z_{i_2}
\end{equation}

\begin{center} 
\includegraphics[width=3.88648in,height=4.0933in,alt={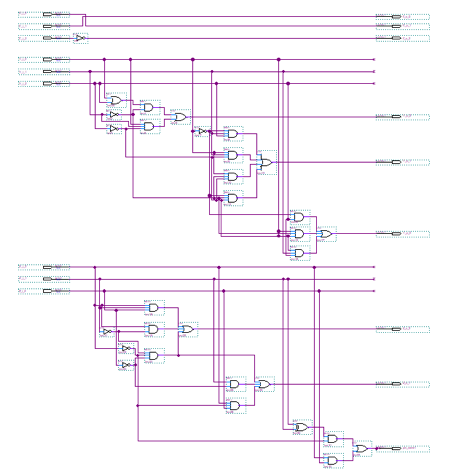}]{digital.png}
\end{center}

The requirements of the digital flag processor module scale
exponentially with the number of modelled qubits since two bits are
required per observable basis state. Therefore, an alternative analog
implementation of the flag processor will also be discussed. But more on
that later.

The more challenging subsystems to implement will maintain the spectrum
of square root probability shared between a qubit\RL{'}s observable
basis states. How this is accomplished will affect how well the outcomes
of measurements in our scheme will match those expected of a true
quantum computer. The probability dimension will be left to an analog
electrical module, which will encode the complete (square root)
probabilistic state into two phasors.

Keeping in mind the goal that at least as much must be knowable about an
emulated quantum state as is expected to be measurable in a theoretical
quantum computing system, let us define a measurable (not in the quantum
sense, but in the classical sense) value (this is not the typical
quantum observable, but rather again something that we can classically
measure about our system) that will enable the description of qubit
state information with a pair of analog waveforms.

Let us consider each of the first and second subsystems separately. Let
each subsystem be responsible for maintaining the magnitude of one term
in each of the coefficients of each observable basis state. Then the
first subsystem might be initialized to maintain the real terms in each
of the coefficients of

each observable qubit state, and the second subsystem might be
initialized to maintain the imaginary terms. Let each of the first two
subsystems be implemented using analog electronics, and be electrically
identical analog modules.

\textbf{Modelling Square Root Probability
Space}

Since an analog module is concerned only with square root probability,
and not sign or phase at this point, its states can be conceptualized as
points within simplexes. This representation will be able to represent
pure quantum states only. It is easy to see that the number of
computational basis states, and the number of vertices on the
corresponding simplex is \(2^{Q}\), where Q is the number of qubits.

Next, we will introduce a mapping from the 3D space of points in a
simplex to a lower dimensional space that can uniquely represent these
quantum states in a way we can classically measure. We will call this
measurable value the Higher Dimensional Observable (HDO) since it is a
proxy for us to measure information in a higher dimensional space.

Consider the following simplex for a two qubit system. If all of the
coefficients in the quantum state are fully real, or fully complex, then
inside the square roots will be probabilities related to one another by
a geometry like this simplex.

\begin{center} 
\includegraphics[width=2.61734in,height=1.83863in,alt={2\_q\_simplex.png}]{2\_q\_simplex.png}
\end{center}

If a coefficient is allowed to be complex, the real coefficients will be
bounded due to the rule that all coefficients' absolute values squared
sum to 1. Each coefficient that is allowed to be complex adds a squared
term to the final sum. i.e. for a 1-qubit state with two complex
coefficients:

\[|il + m|^{2} + |ij + k|^{2} = 1\]

\[\sqrt{l^{2} + m^{2}}\sqrt{l^{2} + m^{2}} + \sqrt{j^{2} + k^{2}}\sqrt{j^{2} + k^{2}} = 1\]

\[l^{2} + m^{2} + j^{2} + k^{2} = 1\]

You can imagine that if we still wanted to represent the real parts of
the coefficients by a simplex that the result of allowing some
coefficients to be partially complex would be like translating the
vertexes in the real simplex towards the centre of the simplex,
according to the bound introduced by these complex values. This would
generalize. In the end, to represent the complex parts of the
coefficients by one simplex, and the real parts by another, the
distances of points from the centre between simplexes would be
correlated by a bound that is less than the extent of the full-size
simplex. Therefore, we can say that any valid pair of points will be
within the original simplexes of maximum volume. Our task is then to map
the space of the original 2 simplexes to the HDO.

If the HDO is to have a unique amplitude for each combination of weights
(or strengths) with respect to each simplex vertex, then distance from
the uniform superposition state along each edge must not be treated
equal. Rather, let us choose arbitrarily that nearness to the maximally
mixed state with respect to \(|0^{\otimes Q}\rangle\) contributes the
most significantly to the strength. Let nearness with respect to
\(|1^{\otimes Q}\rangle\) contribute the least. Let each intermediate
state contribute something between these, with their contribution amount
ordered according to their binary sequence. Then continuously measuring
the strength of the HDO in a system where the state moves gradually from
\(|0^{\otimes Q}\rangle\) towards the uniform superposition state might
yield a curve something like the following.

\begin{center} 
\includegraphics[width=2.79933in,height=1.04869in]{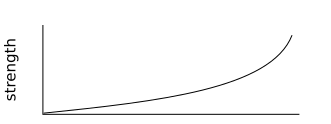}
\end{center}

Continuously measuring the strength of the HDO in a system where the
state moves gradually from \(|1^{\otimes Q}\rangle\) towards the uniform
superposition state would yield the following shape.

\begin{center} 
\includegraphics[width=2.83647in,height=1.00111in]{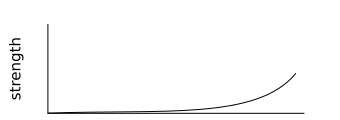}
\end{center}

The nature of the information being encoded suggests that a quantum
harmonic oscillator (QHO) may be an ideal manifestation of the HDO.

At the core of any quantum simulation or emulation is the idea of
mapping quantum information to classical information. At its simplest,
we need each modellable quantum state to be mapped to some classical
state of our classical emulation device. Let \(a \in Z\) be an integer
variable. It will be the goal for this value to be measured or
determined with minimal calculation and minimal deduction. This variable
\(a\) will be used to give each individual modellable quantum state an
identification number.

The HDO must have \(n = 2^{Q}\) observable states, which may be
considered analogous to the energy eigenstates of the quantum harmonic
oscillator. Its observable states must be able to be in superposition,
like the states of the uncollapsed wave function of the QHO. We want to
be able to identify and move between each state in a hierarchy described
by \(a\), which is reminiscent of the raising and lowering operators.

However, a QHO is not ideal for our application for several reasons.
First, we want to be able to identify the precise quantum states, and a
QHO\RL{'}s wave function would collapse to an observable energy state
upon being measured, making this impossible. Second, a perfect quantum
harmonic oscillator is difficult and costly to control.

We will now endeavour to implement the HDO using electronics in a way
that is inspired by the quantum harmonic oscillator but is also
optimally efficient in terms of hardware implementation. The design
should also ensure that it is simple to model the application of the
Clifford matrices to the underlying modelled lower dimensional quantum
system.

First, let the HDO be manifested as a frequency. Let the strength of the
HDO be equivalent to its probability of being consistently measured by a
frequency measurement device. We will now define a hierarchy of
frequencies that accomplish the actualization of this.

Let \(a\) be a measurable integer identifier of a modellable wave
packet. Consider the parabaloid
\(z_{g}(x,y) = gy^{2} + y + gx^{2} + x\). If we were to fix \(y = 0\),
then this would be a parabola with a form loosely analogous to the
classical equation of potential energy in a harmonic oscillator. Imagine
that \(g\) might equal \(\frac{m\omega}{2}^{2}\) and the variables \(x\)
and \(y\) are like position and momentum \(x\) and \(p\) respectively.

\[E = \frac{m\omega}{2}^{2}x^{2} + \frac{1}{2m}p^{2}\]

\[E_{pot} = V(x) = \frac{m\omega}{2}^{2}x^{2}\]

\textbf{The Fundamental P-Spectrum Parabaloid $z_{1}(x,y)$}

Let \(g = 1\). Let \(b \in \mathbb{R}\) be a real number, and the
following set of spheres to be defined in \(\mathbb{R}^{3}\).

\(\{ \cup_{a = 1}^{\infty}s_{a}:(x + \frac{1}{2})^{2} + (y + \frac{1}{2})^{2} + ((z - z_{1}(a,b))^{2} = (a + \frac{1}{2})^{2} + (b + \frac{1}{2})^{2} \}\)

\begin{center} 
\includegraphics[width=3.56202in,height=3.70724in]{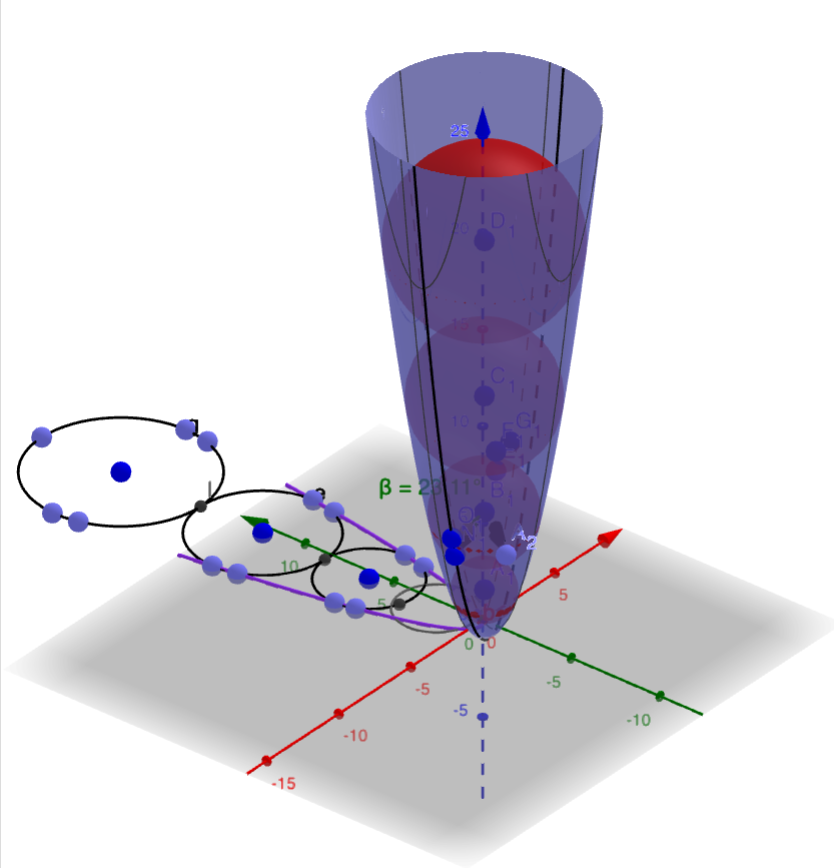}
\end{center}

If \(b\) is left as a free parameter for now, this set of relations is a
set of spheres that are stacked in the shape of a parabaloid such that
the minimum point in the \(z\) axis of each sphere is equal to the
maximum of the preceding sphere, and each sphere is centred in the
parabaloid such that it intersects the parabaloid in a circle, on a
plane parallel to the \(xy\) axis.

\begin{center} 
\includegraphics[width=2.30656in,height=3.14472in]{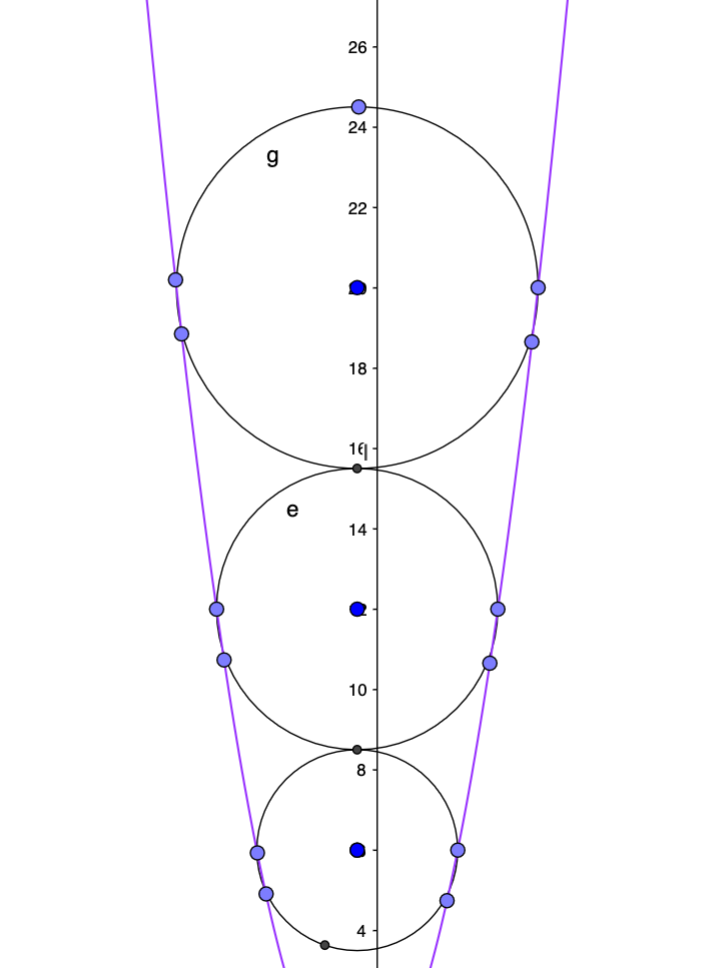}
\end{center}

Then the intersubsection of each circle with \(z_{1}\) will lie on the
plane \(z = z_{1}(a,b)\). Teh values of $x$ for which these intersubsections
occur can be found by solving the following equation.

\[\pm\sqrt{(a + \frac{1}{2})^{2} - (x + \frac{1}{2})^{2}} + a^{2} + a - x^{2} - x = 0\]

However, there will be another, second circular intersubsection of each sphere with
the parabaloid nearby. For each \(s_{a}\) there are four solutions to its intersubsection
equation. Two of the points of intersubsection will lie on the plane
\(z = z_{1}(a,b)\). Let this pair of points be named the
\RL{``}reference points''. The \RL{``}reference vectors'' between the
center of a slice of a sphere and each of its reference points will also lie in the
plane z = z1(a, b) and these vectors will have equal magnitudes.

A sphere\RL{'}s other two points of intersubsection, the \RL{``}data
points'', will each be related to one of the sphere\RL{'}s reference
points. Let a \RL{``}data vector'' be the vector between a data point
and its sphere\RL{'}s center. Let a \RL{``}reference vector'' be the
vector between a reference point and its sphere\RL{'}s center. Let each
data point be related to a reference point such that the magnitude of
the angular displacement of each data vector from its reference vector
is the same. Let this angular displacement be \(\phi\). Then, each
sphere will have a unique single associated angle \(\phi\). This
achieves a mapping from each integer a to a unique \(\phi\). \(\phi\),
of course, could be representable by a single phasor coming from a
single signal
source.

Note the similarity of our \(\phi(a)\) to the potential energy of the
QHO V(x). As the quantized real parameter a increases, the frequency
\(\phi(a)\) will increase as well. \(a\) can be visualized as an integer
in the x-axis. Consequently, a diagram of the energy-wise lowest-lying
solutions of the Schrödinger equation of the QHO also provides a
relatively accurate depiction of our \(\phi(a)\).

\(\phi(a)\) is fundamentally encoded into the \RL{``}data vector'' that
determines its frequency. Given a single data vector, the corresponding
values of \(a\) and \(b\) can be determined. We adopt a coordinate
system with an origin at the center of \(s_{a}\) similar to that of the
Bloch sphere to describe the data vector, where the angle measured from
the x axis is denoted \(\theta\).

Let \(b = f_{2}(r,\theta)\). Let $\theta$ be the angle between two vectors. Let
the first vector be defined by two points: a sphere\RL{'}s centre and
its intersubsection with the plane \(z = z_{1}(a,0)\). Let the second
also be defined by two points: a sphere\RL{'}s centre and its
intersubsection with the plane \(z = z_{1}(a,b)\). Let r be the
magnitude of a data vector. Then we have

\[b = r \cdot sin(\theta)\]

Let \(a = f_{1}(r,\phi)\). This \(f_{1}\) is simply a trigonometric
transformation. Let \(z_{b=0}\) be the parabolic intersubsection of the
plane \(y = 0\) with the parabaloid \(z_{1}\). Let \(\frac{dz_{b=0}}{dx}\)
be the slope in \(x\) of \(z_{b=0}\) at a point approaching a reference
point on the sphere \(s_{a}\). Then we have

\[f_{1}(r,\phi) = \frac{1}{2} ( \frac{r sin(\phi)}{csin(\frac{\pi}{2} - sin^{-1}(\frac{rsin(\phi)}{c}))} - 1 )\]

\[c = \sqrt{2r^{2} - 2r^{2}cos(\phi)}\]

\[dx = csin(\frac{\pi}{2} - sin^{-1}(\frac{r sin(\phi)}{c}))\]

\[dz_{b=0} = r sin(\phi)\]

\[\frac{dz_{b=0}}{dx} = 2x + 1 = 2a + 1\]

\(f_{2}^{- 1}\) is trivial to define:

\[\theta = f_{2}^{- 1}(r, b) = sin^{- 1}(\frac{b}{r})\]

It is simple to define \(\phi\) as a function \(f_{1}^{- 1}\) of \(a\)
and \(b\) as well. Let \(T_{d}(a,b)\) denote the data vector of a sphere
chosen such that \(\phi\) is positive. Let \(T_{r}(a,b)\) denote the
matching reference vector. Then,

\[\phi = f_{1}^{- 1}(a,b) = cos^{-1}(\frac{\vec{T_{r}(a,b)} \cdot \vec{T_{d}(a,b)}}{|\vec{T_{r}(a,b)}| \cdot |\vec{T_{d}(a,b)}|})\]

A graph of \(\phi(a,1)\) iteratively computed using python yields the
following.

\begin{center} 
\includegraphics[width=4.55878in,height=3.00191in,alt={a\_1.png}]{a\_1.png}
\end{center}

\hspace{-15mm}
\includegraphics[width=6.02505in,height=4.92467in,alt={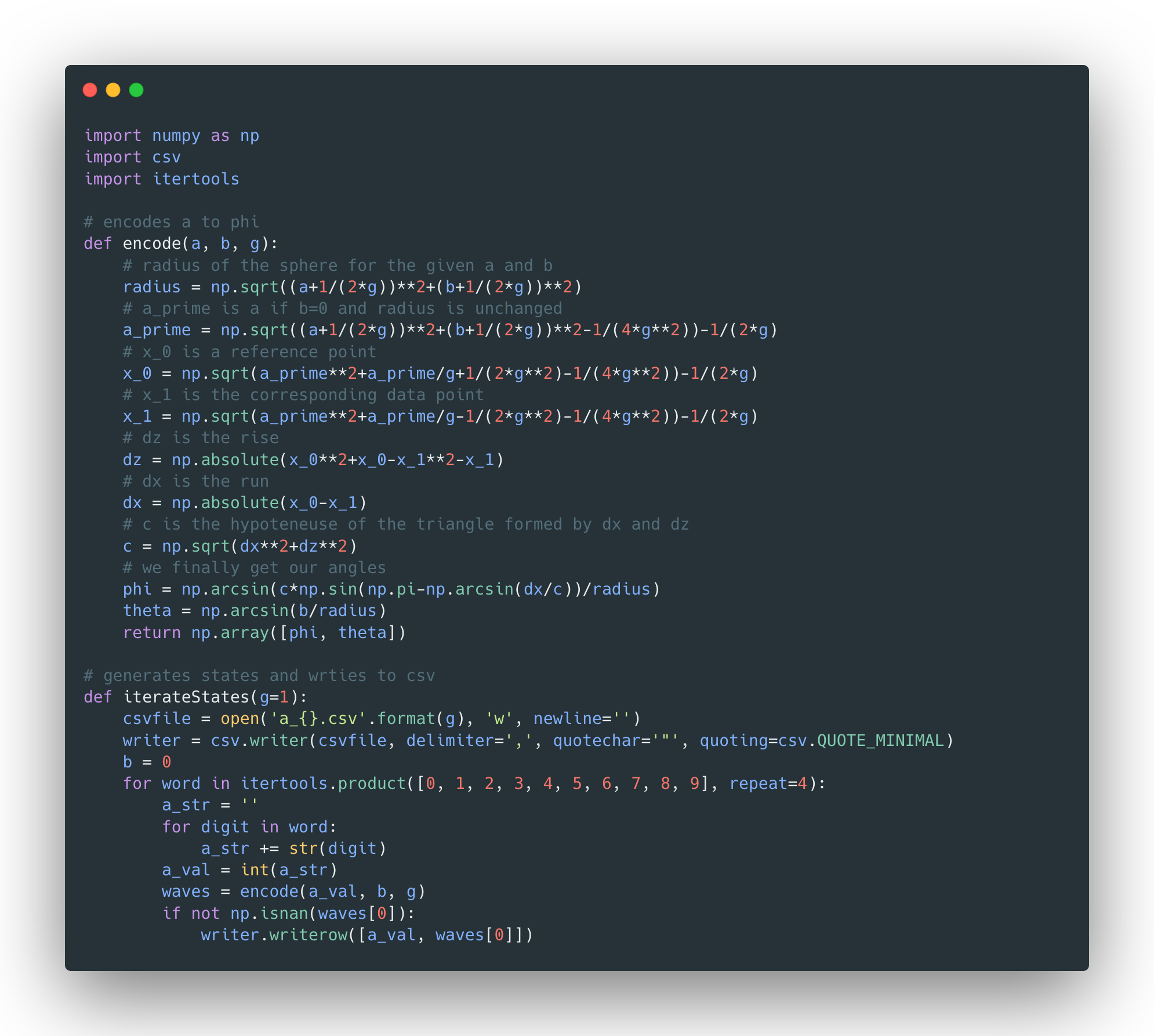}]{carbon28.png}

\textbf{The P-Spectrum Parabaloid Family $z_{g}(x,y)$}

If we return to our initial assumption that \(g = 1\), we can see that
removing this assumption yields a family of functions. Note that after a
certain lower limit in \(a\), the individual functions do not overlap.
The graph for \(0 < g \in \mathbb{Z} \leq 6\) is
provided.

\begin{center} 
\includegraphics[width=4.59073in,height=2.68885in,alt={a\_n.png}]{a\_n.png}
\end{center}

The iterative calculation of the elements of these functions
demonstrates how the functions might alternatively be interpreted as
sequences. This sequential interpretation can help in the categorization
of the space of its elements.

Each of these sequences is a Cauchy sequence. A sequence is Cauchy if
\(\exists\delta > 0:\exists n\epsilon\mathbb{N}:\forall j,k > n||\phi_{j} - \phi_{k}|| < \delta\),
meaning that as the sequence progresses, its elements become arbitrarily
close to one another.

See that a hierarchy of frequencies \(\phi(a)\) has been described such
that if one knows the magnitude of \(\phi\), the value of \(g\) will
also be distinguishable, since the curves \(\phi(a)\) do not overlap.
Knowing both \(g\) and \(\phi\), one may use \(f_{1}\), a simple
trigonometric process, to deduce the value of \(a\).

\hspace{-13mm}
\includegraphics[height=7in]{fig\_2\_8.png}

\hspace{-13mm}
\includegraphics[height=7in]{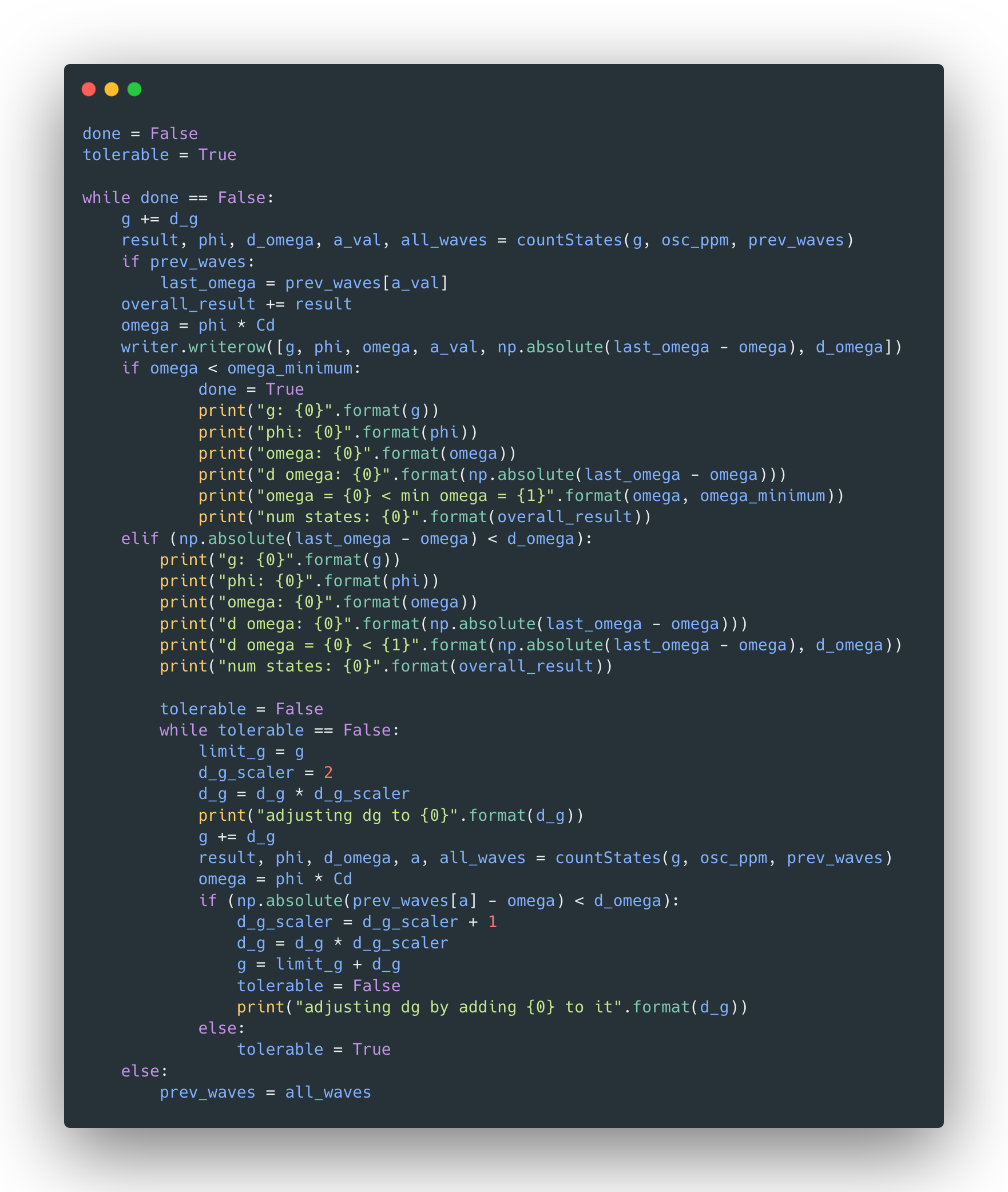}

If we let \(g\) take any real value, then the graph of \(\phi(a)\) becomes
a field. The \RL{``}useful'' elements of such a field
might be determined by introducing an angular resolution d$\omega$ in \(\phi\),
and a maximum value for \(\phi\). The number of \RL{``}useful'' curves
in the vector field can be found by applying the requirement that the
difference in \(\phi\) between any adjacent points in the field is at
least d$\omega$. We can find the set of curves for which the smallest
difference between adjacent points is d$\omega$ and this difference occurs near
the minimum value of \(\phi\).

\textbf{Practical Constraints on d$\omega$ and $\phi$}

High frequency industrial electrical oscillators have ranges that reach
into the tens of Gigahertz. For example, the Axtal AXPLT2500 \cite{axtal_axtal_2016} is a phase
lock crystal oscillator product whose maximum frequency output can reach
12 GHz with a frequency stability $\pm$3.2 ppm depending on operating
conditions age, and other factors. The drawbacks of using such a device
are its physical footprint and its cost.

On the other hand, there are high frequency oscillators with ranges in
the GHz that are available in common integrated circuit chip packages,
such as the \$36.10 Abracon LLC product AX7MAF1-2100.0000C \cite{abracon_2019} which outputs
a maximum stable frequency of 2.1GHz at $\pm$50ppm.

These devices each represent their respective families of devices: the
Axtal product being a part of the family of high end, low noise
industrial oscillator products, and the AX7MAF1 being a part of the
family of small footprint, highly embeddable products. The constraints
introduced by the AX7MAF1 will be analyzed since the goal is to create a
cost-effective and low-profile technology. However, the capacity of
available high-end industrial tools should also be kept in mind.

In the case that one of these oscillators is used as a frequency source
for \(\phi\), d$\omega$ becomes a function of the oscillator's stability
rating so and of \(\phi\) itself.

\[d\omega(s_{o},\phi) = s_{o}\phi\]

We will select values for $a$ and $g$ such that the difference between
\(\phi(a)\) and \(\phi(a - 1)\) is approximately d$\omega$ on the first curve
that contains a value of \(\phi\) that overlaps with the minimum
frequency. The following code iterates through the states in a given
range, determining the number of states for each curve that can be
represented within tolerance. Two device specific parameters for the
procedure are the stability and the minimum output frequency.

Let the highest tolerable value of \(\phi\) calculated by the program be
called the ``base frequency tolerance limit'' \(L_{b}\). In order to
stretch the vector field to fill the acceptable range of operation with
usable curves, we introduce a device specific scaling coefficient
\(C_{d}\) such that \(C_{d} \cdot L_{b} = \omega_{max}\), where
\(\omega_{max}\) is the device's maximum output frequency. Then
we define the relationship between $\omega$ (the device's frequency) and
\(\phi\) (the angle between reference and data vectors) to be
\(\phi = \frac{\omega}{C_{d}}\).

The program can be adjusted to take this into account, spreading the
curves throughout the frequency space of the device. This will yield a
higher achievable number of curves, represented in the program as the
resolution in \(g\).

Running the program using the device parameters for the AX7MAF1
oscillator yields the following range of curves within acceptable
tolerance. ``Distance'' in this chart refers to the distance between the
values of \(\phi\) for adjacent curves at their closest points before 
the tolerance limit is overcome.

\begin{center}
    \includegraphics[width=5.12983in,height=3.54165in,alt={abracon\_tolerance\_range.png}]{abracon\_tolerance\_range.png}
\end{center}

Regarding the ``jumps'' in the graph, consider the following. We are just multiplying by 2 and then 3 each inner loop 
until we find a tolerable state, and sometimes overshooting. Note that this is not a bug. Rather, the jumps in the graph 
are explained by the fact that we sometimes need to go up by more than $1 d_g$ to reach a distinguishable state, in which 
case omega - last\_omega is bigger than it was the previous distinguishable iteration of the outer loop. Overshooting means 
we may underestimate the number of distinguishable states, not the other way around. Some larger jumps have been removed to show 
the trend.

Running the program with the scaling factor \(C_{d}\) and a step of
\(dg = 0.000001\) demonstrates that 4581926 distinct, perfectly
distinguishable values of $a$ could be encoded into frequencies; orders of
magnitude greater than before the scaling factor was introduced. This
demonstrates the effect of the device specific minimum / maximum
frequencies.

\begin{center} 
    \includegraphics[width=4.41763in,height=2.26975in,alt={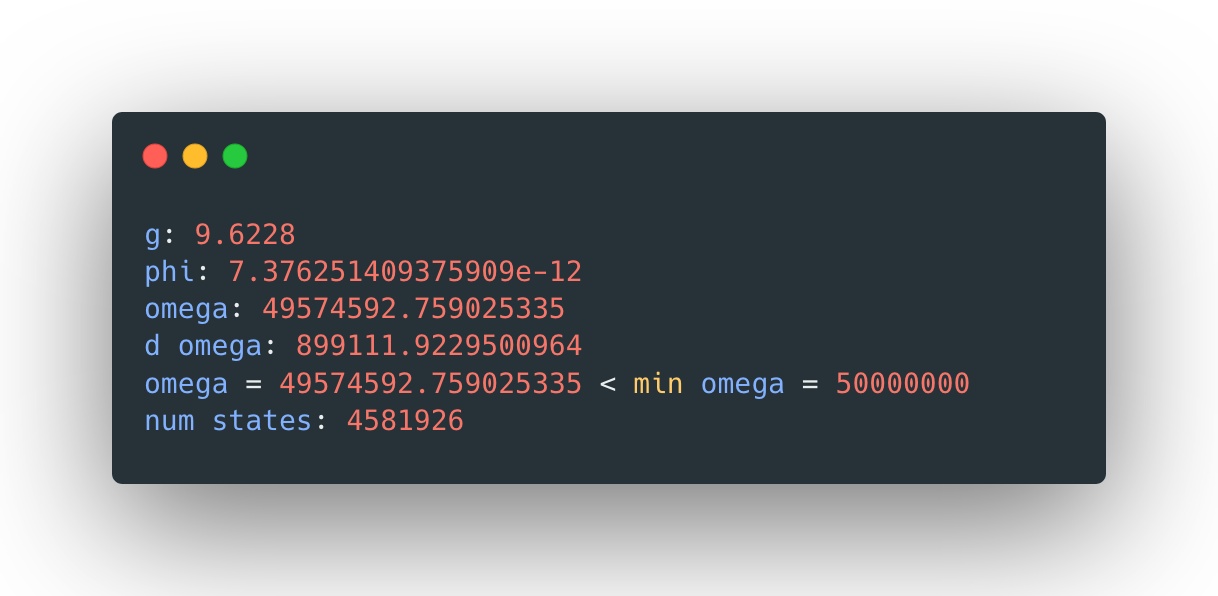}]{carbon34.png}
\end{center}

This number is important since it represents the values of \(a\) that
can be encoded with complete certainty. The goal of this state
representation scheme, however, is to represent probabilistic
information in an analog signal. Therefore, uncertainty must be
introduced in a controlled manner.

\textbf{Probabilistic Measurements}

What we have done so far is to map a set of quadratic relationships to a
similar set of Cauchy sequences of angular values. We then mapped each
of these sequences to a set of practically realizable frequencies. The
convenience of this approach now becomes evident. Since a quadratic
relationship was chosen as the starting point, the frequency encodings
we have described are also each in a sequence where the difference
between adjacent sequence terms changes linearly. Since the difference
between adjacent terms in a sequence corresponding to a value of \(g\)
is equivalent to the difference between two frequencies that are
adjacent in the $\omega$ vector field and we have guaranteed that any of
4581926 terms are perfectly distinguishable within the stability
constraints of a device, we can express the probability of measuring a
particular value of \(a\) in terms of only the difference
\(\omega(\phi(a)) - \omega(\phi(a - 1))\) and the resolution of a
frequency measurement device.

Let \(\sigma(a,g)\) be the variance in measured values for the value of
\(\phi(a,g)\),
and\(\delta(a,g) = \omega(\phi(a,g)) - \omega(\phi(a - 1,g))\) be the
angular resolution of the medium maintaining the angle \(\phi\). Then
the probability of measuring values of \(a\) and \(g\) can be expressed
as the following.

\[
    p(a,g) = \frac{\delta(a,g)}{\sigma(a,g)}
\]

Let this probability be equivalent to the strength of the HDO. Note that
a value of \(a\) with a particular probability of being measured is now
mappable to a specific encoding frequency. Recall that the original
purpose of \(a\) was to enumerate modellable quantum states. The next
task is to create a mapping between each value of \(a\) and its
corresponding quantum state such that the probability of measuring a
value of \(a\) with a measurement device is equal to the strength of the
HDO.

Recall that the greatest contributor to the strength of the HDO was
chosen to be the distance from the maximally mixed state with respect to
the \(|0^{\otimes Q}\rangle\) state. Therefore it is logical to assign
the \(g = dg\) curve to correspond to states that lie exactly between
the uniform superposition state and the state \(|0^{\otimes Q}\rangle\).

The rest of the states will be related to curves evenly spaced
throughout the frequency range of the device used to maintain \(\phi\).
For example, using the AX7MAF1 to model a two qubit system would yield
the following ``chief g curves''. The spectrum will be made to wrap back
to the first state so as to

better map the space inside the simplex to its curves.

{\def\LTcaptype{none} 
\begin{longtable}[]{@{}
  >{\raggedright\arraybackslash}p{(\linewidth - 2\tabcolsep) * \real{0.4813}}
  >{\raggedright\arraybackslash}p{(\linewidth - 2\tabcolsep) * \real{0.4813}}@{}}
\toprule\noalign{}
\begin{minipage}[b]{\linewidth}\raggedright
\textbf{Chief g curve}
\end{minipage} & \begin{minipage}[b]{\linewidth}\raggedright
\textbf{State}
\end{minipage} \\
\midrule\noalign{}
\endhead
\bottomrule\noalign{}
\endlastfoot
dg & \textbar00\textgreater{} \\
6 & \textbar01\textgreater{} \\
12 & \textbar10\textgreater{} \\
18 & \textbar11\textgreater{} \\
24 & \textbar00\textgreater{} \\
\end{longtable}
}

The system is also capable of representing states do not lie on any of
the chief curves. Since dg used in the simulation generating these
curves was 0.000001, we know that we could use many curves
between each of the four in the table above to represent such states.

See that the effect of the application of an X gate on a qubit would be
to move a state in its simplex by reflecting it across a line that
passes through the centre and is orthogonal with an edge representing a
transition between two states with the operand qubit's value
changing and the rest not changing, for each such edge that exists. We
will partition the curves between the chief curves with the goal of
keeping this as simple as possible in practice.

Let the intermediate curves between each chief curve be evenly divided
into a number of groups called ``surface groups''. Let the set of curves
related most closely to a vertex be called a ``vertex group''.

In the illustration below, each vertex group is given a colour, and each
surface group is labelled with a number indicating its order within its
vertex group.

\begin{center} 
  \includegraphics[width=4.92939in,height=3.71164in,alt={surface\_groups.png}]{surface\_groups.png}
\end{center}

To envision the way that the simplexes represent probabilities, the
illustration above is most helpful. However, to visualize the overlay of
\(g\) curves properly over the space of square root probability and have
things line up, it is better to picture tetrahedral spheres. The
topology is identical to that of the simplex, only the edges are made to
curve so that the overall shape is spherical. This will help us to
understand how an X operation can be
implemented.

\begin{center} 
\includegraphics[width=2.2in,alt={tetrahedral\_sphere.png}]{tetrahedral\_sphere.png}
\end{center}

\textbf{Conceptual Introduction to Possible Operators}

See that applying X to the last qubit will move a state from one surface
group to a different surface group. It can be implemented by the
addition or subtraction of a constant from \(g\). Let's call this
constant \(\frac{V_{ss}}{4}\) for now. The reason for this will become
apparent later. We will build from a simple example to more complex
examples.

\begin{center} 
\includegraphics[width=2.55779in,height=1.40485in,alt={X\_g\_example\_1.png}]{X\_g\_example\_1.png}
\end{center}

For the claim that X can be implemented by addition or subtraction of a
constant from \(g\), the values of \(g\) that correspond to the
following similarly marked pairs of quantum states must be equidistant.
Meaning, the differences between the similarly marked values are equal
to our constant.

\begin{center} 
\includegraphics[width=2.2in,alt={X\_g\_example\_1\_sphere.png}]{X\_g\_example\_1\_sphere.png}
\end{center}

See that these are states along an edge. Notice how the first and second
half of the range of \(g\) depicted each represent a different half of
the edge, with the direction along the edge inverted with respect to the
other half.

\begin{center} 
\includegraphics[width=3.5103in,height=1.52051in,alt={X\_g\_example\_2.png}]{X\_g\_example\_2.png}
\end{center}

Next, consider a surface state. To do this we'll add a variable strength
with respect to \(|01\rangle\). The overall state would be normalized
depending on the value \(x\), so that the proportionality between
the \(|00\rangle,|01\rangle\) terms would be preserved.

\begin{center} 
\includegraphics[width=2.2in,alt={X\_g\_example\_2\_sphere.png}]{X\_g\_example\_2\_sphere.png}
\end{center}

See that the same principle of bifurcation applies, this time so that we
invert the direction with respect to the \(|00\rangle,|01\rangle\) edge
and the direction with which we translate the states by \(x\) away from
the edge.

\begin{center} 
\includegraphics[width=4.8in,alt={X\_g\_example\_3.png}]{X\_g\_example\_3.png}
\end{center}

These transformations have thus far only involved \(g\) since all the
states have been equidistant from the centre of the tetrahedral sphere.
However, in order to perform a general X gate, we need to also consider
the final strength with respect to
\(|11\rangle\).

\begin{center} 
\includegraphics[width=4.34348in,height=1.58908in,alt={X\_g\_example\_3\_sphere.png}]{X\_g\_example\_3\_sphere.png}
\end{center}

In the previous example, we drew our marks on two adjacent surfaces. The
effect of adding this dimension will be to translate the surfaces on
which we draw our marks towards opposite vertexes. This is how the
bifurcation presents in this final example. The effect is that our
states lie on surfaces within the tetrahedral sphere instead of on it.
These inner surfaces have less surface area than the outer ones did,
this shrinking proportional to \(y\).

Note how \(y\) and therefore the distances of the inner surfaces from
the centre of the tetrahedral sphere do not change before and after the
application of the X gate. Therefore, \(a\) remains unaffected even in
this most general of the examples and only \(g\) is changed.

In the end we want to represent all of these states. We would like to
somehow interleave them, roughly illustrated by the following.

\begin{center} 
\includegraphics[width=4.2in,alt={X\_g\_example\_4.png}]{X\_g\_example\_4.png}
\end{center}

This is a grainy mapping of 3D space to 2D. We iterate not over the
values of the coefficients in the quantum state, but rather over their
relative proportions to one another.

We previously defined vertex groups. A vertex group is bound by
comparison to the coefficients of the 3 kets not labelling the chief
vertex in question. The surface groups are defined by the coefficients
corresponding to two non-chief vertices being greater than the final
coefficient. This outlines an approach to our mapping for us.

We would divide the full range of \(g\) into the 4 vertex groups. This
would mean that in the quarter of the spectrum corresponding to a
vertex, the strength w.r.t. that vertex is the highest of all the
strengths. Within each vertex groups we would always have three other
strengths, which we will call \(x,y,z\). We will assert that
\(x = \delta_{1} < \delta_{0},y = \delta_{2} < \delta_{1},z = \delta_{3} < \delta_{2}\)
for varying \(\delta_{i}\) and \(\delta_{0}\) associated with the chief
vertex. What is important is that these varying values \(\delta_{i}\) of
\(x,y,z\) always vary together in the same way with respect to their
ordering, to iterate over all of the relative strengths of the vertexes
within the appropriate bounds that can be mapped with a graininess
determined by the number of \(g\) curves we can represent.

The vertices that the \(x,y,z\) are associated with will change
depending on the sixth of the vertex group that we are in. i.e. in the
first sixth of the first vertex group we would have
\(x_{01} = \delta_{1} < \delta_{0},\)\(y_{10} = \delta_{2} < \delta_{1},z_{11} = \delta_{3} < \delta_{2}\)
and in the second
\(x_{01} = \delta_{1} < \delta_{0},\)\(y_{11} = \delta_{2} < \delta_{1},z_{10} = \delta_{3} < \delta_{2}\).
As we iterate through the sixths of the vertex group's \(g\) range, we
would permute the subscripts of the \(x,y,z\). This can be done
consistently across all the vertex groups so that each surface group is
fully represented in its vertex group and the addition or subtraction of
the same constant from \(g\) can implement the Pauli X gate on the last
qubit.

In the two qubit system, a X gate or similar operation can be easily
recognized by simply partitioning the entire set of possible modellable
our states into 24 such regions. For more complex operations, the number
of partitions required might increase. But this serves as a template.

It will be clear to the reader at this point that the scaling and
complexity of representing quantum information has not been side-stepped
by any sort of black magic. It is present in this mapping. Instead, the
limiting factor has been changed from the scaling of algorithms in time
(in terms of runtime on a CPU or in a way that is limited by bandwidth
as described earlier) and space (such as classical memory), to the time
of analog modules' settling times and space in terms of the precision of
available analog modules. These aspects will be addressed concretely by
a set of hybrid digital / analog circuits that can perform Clifford
gates and measurements for us, as well as encode \(a\) and \(g\) to
\(\phi\) in order to control an oscillator.

\subsection{Dual Oscillator Electrical Implementation
Design}\label{dual-oscillator-electrical-implementation-design}

\textbf{High Speed Analog Module Design}

A high speed analog module (HSAM) will be responsible for converting
\(a,g\) to \(\phi\). Below we sketch out this circuit. See that
additional required input signals are:
\(V_{\frac{1}{2}},\omega,V_{x}\) and t.

Use of the module would consist of a few sequential steps:

1. Setting the voltage \(V_{\frac{1}{2}}\) one half of the value of 1 in
the desired numerical precision of the analog calculation module. This
should be scaled to account for the voltage range constraints of the
voltage source used to power the module.

2. Setting the voltage at the input \(V_{g}\) and \(V_{a}\) equal to
\(2gV_{\frac{1}{2}}\) and \(2aV_{\frac{1}{2}}\) respectively.

3. Setting \(V_{x}\) to $V_{a}$ and
sweeping the voltage range to \(V_{x} = 2(a - 1)V_{\frac{1}{2}}\). 
It is necessary to include the upper boundary of this range in the sweep.

Note that when we are adding scaled values like i.e. $0.1 + 0.5$ to represent
$1 + 5$, we get the desired outcome i.e. $0.6$ that represents $6$. However,
when multiplying or dividing, things work differently. Consider that
$\frac{1}{0.5}$ does not represent $\frac{10}{5}$ since $\frac{1}{0.5} = 2 
\neq 0.2$. Therefore, we need to re-scale the output from certain types of 
operations. This re-scaling is excluded in the following diagram, but should 
be considered implicit.

The module expects an accurate clock pulse to be present at the input
labelled t, which should represent 1 unit of time. Increasing the clock
pulse will increase the output frequency proportionally, and can
therefore provide a mechanism by which an output frequency
representation of $\omega$ can be scaled to meet any frequency bandwidth
constraints. However, any scaling must be accounted for when $\omega$ is
measured, or the value of \(\phi\) will be misinterpreted.

The module also expects a frequency source to be connected. This is
shown in the diagram as an AC voltage source. The frequency source has
one input and one output. The output is expected to be the frequency $\omega$
(or a scaled version of $\omega$) and the input will be a control voltage
proportional to the magnitude of the frequency shift that is required to
encode the quantum state specified by \(g\)

and \(a\) into the frequency. The adjustment of the oscillation that is
effected by the control voltage is a traditional closed loop control
systems problem.

\textbf{High Speed Analog Module Efficiency}

The speed of any quantum operation applied by this module is related to:

\begin{itemize}
  \item The settling time of the operational amplifiers in the circuit.
  \item The accuracy affected by the value of \(V_{d}\).
\end{itemize}

Settling time is the elapsed time from the application of an ideal
instantaneous input voltage to the time at which the output has entered
and remained within a specified error band. The settling time can be
determined form the propagation delay or slew rate, which can be found
in all op amps\RL{'} data sheets.

Let us assume we were to use a modern op amp, the LMH3401. LMH3401 op
amps \cite{texas_2014} have a slew rate of \(18000\frac{V}{\mu s}\) . Assuming an
operating power range of 10 volts, the total delay due to settling time
will be less than \(\frac{10V}{18000\frac{V}{\mu s}} = 5ns\) per
sequential op amp in the longest chain of op amps. Therefore it will be
in the order of 50 ns.

\begin{center} 
  \includegraphics[height=7.75in,alt={analog\_module\_marked\_up.png}]{analog\_module\_marked\_up.png}
\end{center}

The accuracy affected by the value of \(V_{d}\) can also be bounded. The
value of the voltage \(V_{d}\) is related to the closeness of the data
vector used in the experimental calculation of $\omega$ to the theoretical
magnitude and direction of the vector for the given value of \(a\) and
\(g\). The relationship can be described using
trigonometry.\begin{center} 
\includegraphics[width=2.08954in,height=2.40155in,alt={ratio\_tr\_exp.png}]{ratio\_tr\_exp.png}
\end{center}

\(V_{d}\) is exactly the difference between the magnitude of the
experimental data vector which is created as \(V_{x}\) sweeps, and the
magnitude of the experimental reference vector. The magnitude of the
reference vector is manifested as a voltage itself, and equal to
\((V_{x} + V_{\frac{1}{2}}) \pm V_{d}\).

Let the angle between the true data vector and experimental data vector
be \(\phi'\). Then the magnitude of \(\phi'\) will be zero when they
vectors\RL{'} magnitudes are equal.

Recall that $\vec{T_{d}(a,b)}$ denotes the data vector of a sphere chosen
such that \(\phi\) is positive, and $\vec{T_{r}(a,b)}$ denotes the matching
reference vector.

\[\phi = cos^{-1}(\frac{\vec{T_{r}(a,b)} \cdot \vec{T_{d}(a,b)}}{|\vec{T_{r}(a,b)}| \cdot |\vec{T_{d}(a,b)}|})\]

Then, let us introduce an error term $\Delta$.
Here, the typical "relative error" is $(1 \pm \delta) = \frac{|\vec{T_{d}(a,b)}| \pm \Delta}{|\vec{T_{d}(a,b)}|}$.
We introduce a single relative  $(1 \pm \delta)$ to the vector $\vec{T_{d}(a,b)}$ and also to its magnitude. This is not a simplified
error model, but works exactly since a relative error term introduced in the individual components of a 2-vector (remember b=0) is 
the same as that term introduced to the magnitude of the vector. i.e.

\[
  \sqrt{(x\delta)^2 + (z\delta)^2} = \sqrt{(x^2 + z^2)\delta^2} = \delta \sqrt{x^2 + z^2}
\]

This allows us to multiply and divide by $(|\vec{T_{d}(a,b)}| \pm \Delta)$ and keep the equation balanced.

\[
  cos(\phi) = \frac{\vec{T_{r}(a,b)} \cdot \vec{T_{d}(a,b)} \cdot (1 \pm \delta)}{|\vec{T_{r}(a,b)}||\vec{T_{d}(a,b)}| \cdot (1 \pm \delta)}
\]

\[cos(\phi) = \frac{\vec{T_{r}(a,b)} \cdot \frac{\vec{T_{d}(a,b)}}{|\vec{T_{d}(a,b)}|} \cdot (|\vec{T_{d}(a,b)}| \pm \Delta)}{|\vec{T_{r}(a,b)}| \cdot (|\vec{T_{d}(a,b)}| \pm \Delta)}\]

Let the inverse multiplier

\[Inv_m = \frac{|\vec{T_{r}(a,b)}||\vec{T_{d}(a,b)}|}{\vec{T_{r}(a,b)} \cdot \vec{T_{d}(a,b)}}\]

and

\[h = cos(\phi) Inv_m\]

Simple algebraic manipulation yields the relationship of $\Delta$ with \(\phi\):

\[
  cos(\phi) Inv_m = \frac{|\vec{T_d(a,b)}| \pm \Delta}{|\vec{T_d(a,b)}| \pm \Delta}
\]

\[
  |\vec{T_d(a,b)}| \pm \Delta = \frac{|\vec{T_d(a,b)}| \pm \Delta}{cos(\phi) Inv_m}
\]

\[
  |\vec{T_d(a,b)}| = \frac{|\vec{T_d(a,b)}| \pm \Delta}{cos(\phi) Inv_m} \mp \Delta
\]

\[
  |\vec{T_d(a,b)}| = \frac{|\vec{T_d(a,b)}| + (\pm 1 \mp cos(\phi) Inv_m) \Delta }{cos(\phi) Inv_m} 
\]

\[\Delta = \frac{cos(\phi) Inv_m |\vec{T_d(a,b)}|-|\vec{T_d(a,b)}|}{(\pm 1 \mp cos(\phi) Inv_m)}\]

Then,

\[\Delta = \frac{(h - 1) \cdot |\vec{T_d(a,b)}|}{\pm 1 \mp h}\]

Essentially, \(h\) contributes to the uncertainty with a magnitude of
\(\pm 1\), except for in the region where $\Delta$ → 0.

Therefore, $\Delta$ is at worst \(\pm |\vec{T_{d}(a,b)}|\) for all values
of \(h\):

\[\Delta \leq \pm |\vec{T_{d}(a,b)}|\dot{=} \pm (x + \frac{1}{2})\]

The implication is that the maximum effect of \(V_{d}\) is bounded by
\(x + \frac{1}{2}\). $\Delta$ may cause a discrepancy in the encoded value of
\(a\) proportional to the magnitude of \(a\) itself.

This can be bounded further since any data point will lie on the surface
\(z_{g}(x,y) = gy^{2} + y + gx^{2} + x\). \(x\) is related to \(g\)
and to $a$ such that if \(g\) increases but \(a\) remains the same, \(x\)
will decrease. If \(x\) decreases, so will $\Delta$, until \(x = \Delta = 0\).
On the other hand, $\Delta$ will increase as \(g\) → 0. \(g\) cannot be exactly
0, but \(V_{g}\) might take a value equal to \(2V_{\frac{1}{2}}\), the
representation of a numerical 1, in order to represent closeness to the
\(|0^{\otimes Q}\rangle\) state. Let this be considered close to the
worst case. In this case, \(a = \lceil x\rceil\). We again introduce the
error:

\[a = \lceil x \pm \Delta\rceil = \lceil x \pm (x + \frac{1}{2})\rceil\]

The worst case scenario for this condition would be that \(x = a\) See
that $\Delta$ would then at most cause a discrepancy of \(da_{\Delta} = 2a\).

Finally, we need to take into account that this error only plays a role
in the calculation of data vectors, which is done during the sweep of
\(V_{x}\) from \(V_{x} = V_{a}\) to \(V_{x} = 2(a - 1)V_{\frac{1}{2}}\).

So, the maximum error will actually be limited by the difference
between \(V_{X_{d}}\) (the data vector's value of \(V_{x}\)) and
the starting value in the sweep \(V_{x_{0}}\).

The trade-off with this error is speed. If \(d_{a_{\Delta}}\) is greater
than the difference between \(V_{X_{d}}\) and \(V_{x_{0}}\) , then the
sweep will trigger the rest of the module's calculations
immediately. Otherwise, it will wait until a voltage close enough to the
data vector's value of \(V_{X_{d}}\) has been included in the
sweep.

The speed of any quantum operation applied by the analog module is
directly proportional to:

1. The speed of the voltage sweep from \(V_{x} = V_{a}\) to \(V_{x} = 2(a - 1)V_{\frac{1}{2}}\).

2. The frequency of the scaled time reference introduced by the clock
source t.

These factors both depend on the quality of external equipment. The most
significant is the speed of the voltage sweep. A voltage sweep's
speed is limited only by the ability of the circuit to respond to it.
Therefore, we can use the worst case scenario that the settling time of
each op amp is 5 ns to be the slowest sweep time required.

Since the frequency scaling feature cannot scale a frequency past the
stability limit of the $\omega$ oscillator, it is only potentially useful for
lower frequencies. However, it does have the potential to increase
performance.

The voltage signal output to the oscillator is updated once per unit
time. So, while the speed of the calculation is not affected by the time
scaling feature, the speed at which data is transmitted to the next
device is.

The way that the adjustment voltage for the oscillator is calculated is
based in simple trigonometry. Two vector magnitudes are calculated and
passed to the final segment of the analog module, r and r'.

\begin{center} 
\includegraphics[alt={ratio\_rs.png}]{ratio\_rs.png}
\end{center}

The relationship between r, r' and \(\phi\) is:

\[\frac{r'}{r} = cos(\phi)\]

The signal $\omega$ produced by the oscillator (before being multiplied by
\(C_{d}\)) has frequency \(\phi\) and is therefore \(cos(\phi t)\).

So, for $\omega$ to be correct, the DC voltage value of $\omega$ must equal
\(\frac{r'}{r}\) periodically in time, when t is a multiple of 1 time
unit.

The "frequency control output" will become zero when $\omega$ reaches the correct
new value, as the frequency of the oscillator is adjusted. The initial value of the 
frequency control output after a calculation is a voltage that indicates how far off 
the current frequency is from the next value. It is a classic control system feedback loop.

Therefore, the time required to update the oscillator after a
calculation is ideally equal to 1 time unit, which may be scaled by the clock
source. However, this will be limited by the speed at which the oscillator can modulate
its frequency.

In summary, we have the ideal total time \(t_{tot}\) required for
updating the state to reflect the outcome of a quantum transformation:

\[t_{tot} = 1\text{scalable time unit} + 50ns + (\frac{V_{x_{d}} - \Delta - V_{x_{0}}}{18000 V}) \mu s\]

Here $(V_{x_{d}} - \Delta - V_{x_{0}})$ approximates the voltage range 
that is needed to be swept before the rest of the $50ns$ delay, and finally the 1 scalable time unit, can be incurred.

Note that the complexity of a quantum operation is not significant to
this module, as long as the final result is deﬁned by \(g\) and \(a\).
The complexity of an operation is the responsibility of an external
hardware module. For example, we know that applying an X gate to the
last qubits in a system is simply equivalent to adding or subtracting a
constant. So, the external module for doing this
would simply be an analog arithmetic circuit. This operation's time
of actualization would be equal to the HSAM's \(t_{tot}\) plus the
propagation delay of the external module.

\textbf{Encoding Sign and Phase}\label{encoding-sign-and-phase}

We will now discuss an approach to encode phase and state information
into the same signal as the probability distribution of a 
state's observables. This would replace a digital flag module.

\subsubsection{Bandwidth Limited Approach}

We will now discuss an approach to encode phase and state information
into the same signal as the probability distribution of a 
state's observables. This would replace a digital flag module.

Note that we have not yet taken advantage of one rotational axis in the
parabaloid-based model of the information, and that any rotation in $\theta$
can be interpreted as a rotation of the cartesian frame about the vertical
axis through the center of a sphere, and so not affect the results of the calculations of g or a. It is
an additional degree of freedom that can be used to encode phase and
sign into the signal that represents our mixed qubit state.

Let us assume that every output of the HSAM described thus far is a pure
frequency shift of the original oscillator's frequency and no phase
shifting is induced by this operation. The HSAM described would then
operate perfectly on completely DC voltages. None of the components
within it have significant reactivity. So, we can immediately redefine
the input signal to this module to oscillate with a frequency and the
functionality of the HSAM's internal circuitry will not change. The
only difference is that it will act on the DC component of the signal
being processed rather than the entire signal. The root mean square
(RMS) voltage \cite{CP_320} will then be the component of the signal that emulates the
probability space, where T is the period of the oscillation:

\[V_{RMS} = \sqrt{\frac{1}{T}\int_{0}^{T}V_{max}^{2}cos^{2}(w t)dt}\]

Simplified, the RMS voltage can be roughly calculated from the peak
voltage of an oscillation \cite{CP_320}:

\[V_{RMS} = \frac{1}{\sqrt{2}}V_{max}\]

The control signal that is sent to the AX7MAF1 oscillator will then
oscillate as well. The effect will be for the frequency $\omega$ of the final
output signal to oscillate in time. We can phase shift this order 2 frequency
to encode our sign and phase information.

The number of qubits in the emulated system directly dictate the number
of phase and sign permutations possible. The set of permutations will
have a size equal to \(2^{2^{Q+1}}\) (2 bits per ket), to account for each of the negative
and imaginary flags. The maximum required permutations for one 20 qubit
processor would then be $2^{2^{Q+1}}$.
If flag processing is implemented as an AC signal processor, then the
phase and sign of the emulated quantum basis states' coefficients
will be manifested as oscillations of the frequency output $\omega$ from the
AX7MAF1 high frequency oscillator. These oscillations of $\omega$ will
encode information in their phase. The frequency requirements to encode
sufficient phases will now be bounded.

If we have Q qubits and need to be able to identify
$2^{2^{Q+1}}$ separate phases in the frequency of the output
oscillation, then we need to be able to measure the frequency of the
output quickly enough to have a sampling frequency at which
$2^{2^{Q+1}}$ separate phases can be identified.

In order to digitally measure the frequency we might use a translation
circuit with three stages:

1. Signal smoothing with a capacitor

2. Schmitt trigger conversion to square wave

3. Amplitude fixing with a Zener diode

\begin{center} 
\includegraphics[width=3.59536in,height=1.40214in,alt={freq\_measure.png}]{freq\_measure.png}
\end{center}

The beneﬁt of the phase shift encoding of phase and sign data is that
regardless of the value of the phase shift, it will always take at most
a time equal to the period of the phase shifted oscillation of $\omega$ to
measure. This measurement can be done by simply detecting at which index
into the generated digital signal the high level of a specific length
occurs.

We need to be able to identify $2^{2^{Q+1}}$ separate phases. This means
there will be \(\frac{2^{2^{Q+1}}}{2}\) peaks of the output frequency
per period of the phase shifted output frequency oscillation. 
Here, we are approximating the digitized signal as a series of time slices, any one 
of which is either a peak or a valley of $\omega$. At one of these time slices, if the 
frequency $\omega$ is itself oscillating, the length of a peak will cross a threshold. 
The index of this slice will indicate the magnitude of the phase shift. We need
\(\frac{2^{2^{Q+1}}}{2}\) peaks since the number of possible time slices is approximately
the nubmer of peaks plus the number of valleys, or the number of peaks times 2.

The time it would take to measure all of these peaks is approximately
\(\frac{2^{2^{Q+1}}}{2\omega}\). This is because, at worst, we would need to measure the entire period 
of the second order oscillation, which would be equal to the period of the first order oscillation times
the nubmer of peaks, since there would be an equal number of periods of the first order oscillation
shorter than $\frac{1}{\omega}$ and longer than $\frac{1}{\omega}$, by amounts that would cancel.
This would have a worst case when $\omega$ approaches 50MHz, and a best case when $\omega$ = 2.1Ghz.

The time to measure a quantum system's state coefficients'
phases and signs would then take between
\(\frac{2^{2^{Q+1}}}{2 \cdot 50MHz}\)
and \(\frac{2^{2^{Q+1}}}{2 \cdot 2.1GHz}\) if the entire period were measured. This is intractable for say, 
20 qubits, and is in that sense comparable to the performance of the device proposed in 2016.

However, we can rethink the approach a bit to do better. Measuring all of
these peaks is unnecessary in our system since the only information that
is required to identify a phase shift is the length of the first peak. From it, 
we can determine where in the phase shifted second-order oscillation we are.
Therefore, our worst case need only be close to
\(\frac{1}{\omega_{min}} = \frac{1}{50MHz}\overset{\cdot}{=}20ns\),
and the best case will be
\(\frac{1}{\omega_{max}} = \frac{1}{2.1GHz}\overset{\cdot}{=}0.476ns\).
The limitation then once more becomes how many distinguishable frequency
shifts we can represent using our oscillator, which we have seen before.
We will at least be able to represent enough permutations of the flag bits
for our 2-qubit proof of concept. Beyond small numbers of qubits, the bandwidth
becomes the problem which translates to time. It would take forever to measure
the phase or sign of a state. However, notice that we specifically split the sign
and phase into this separate module, since they never need to be measured! The
probability of observing a basis state is equal to the square of the magnitude of 
its complex coefficient, and the square of the magnitude of a complex number 
is independent of its phase and sign. Since we maintain the imaginary and the real
components in separate subsystems, we always deal with either purely imaginary 
or purely real coefficients at a time. The magniutde of a purely imaginary coefficent $mi$
is just the absolute value of $m \in \mathbb{R}$.

Let the amplitude of the phase shifted output frequency oscillation be
fixed for simplicity, and be chosen to be minimal. The sampling
frequency of our frequency measurement circuit must not interfere with
these best and worst case scenarios. The delay of a Schmitt trigger is
roughly 0.125 ns. The propagation delays of the other discrete
components of the output frequency digitization measurement circuit are
trivial. Since \(0.125ns < \frac{1}{2.1GHz}\overset{\cdot}{=}0.4ns\),
the measurement circuit will perform well in the entire range of output
oscillation as long as the amplitude of the phase shifted output
frequency oscillation is chosen to be less than
\(\frac{1}{0.125ns}\overset{\cdot}{=}8GHz\), which is impossible due to
the limitations of the frequency oscillation device. Before the
amplitude of the phase shifted $\omega$ oscillation is manifested in the output
oscillation itself, it will be computed as an AC amplitude on the
control signal being processed by the HSAM. We will choose a value of
dVg less than the value dg which was used
earlier to estimate the density of g curves that can be emulated in
general. Therefore this oscillation will not interfere with the
properties of g curves that have been determined. This will also have a
minimal effect on the number of encodable value of phi before $\omega$
saturates. This will also help minimize the addition to measurement time
due to the phase and sign information.

The frequency of the phase and sign oscillation of the control signal is
not especially consequential but should be chosen such that the
frequency responses of average and affordable reactive components such
as capacitors can be used to perform meaningful phase shifts.

\subsubsection{Alternative Hybrid Analog / Digital Approach}

In order to overcome the bandwidth limitations of the approach outlined above for larger
system sizes, we can instead use the following hybrid analog / digital approach.

When we perform a quantum computation, we never apply an operation that might involve 
a sign or phase flip to a single ket. Instead, because quantum mechanics is a linear theory,
we apply this operation to each ket individually. This guarantees that when we apply 
an operator $U$ to a ket $k_0 = (-1)^s (i)^p \ket{b_0b_1\dots b_n}$, we also apply it to
$k_1 = (-1)^{s} (i)^{\bar{p}} \ket{b_0b_1\dots b_n}$. If the operation has the effect of multiplying 
the state $\ket{\vec{b}}$ by $i$, then a few things happen. First, the ket $k_0$ (if $p=1$) or $k_1$ (if $p=0$) with an $i$ 
prefactor loses its $i$ prefactor, since $i^2=-1$. Then, the ket $k_1$ (if $p=1$) or $k_0$ (if $p=0$) initially 
\textit{without} an $i$ pre-factor gains an $i$ pre-factor. Finally, we need to update the sign bit
$s$ of the first ket.

Recall that we maintain our amplitudes using two tetrahedral spheres encoded into two separate 
oscillators, one for the kets with $i$ pre-factors, and one for those without. See that a "multiply by i"
operation therefore simply involves swapping the strengths with respect to kets $k_0$ and $k_1$ between
oscillator modules, and updating the sign bit associated with the first.

In terms of sign, we would normally need one sign bit for each ket, which is exponential in the system size.
However, we can reduce this by using a sparse representation. In a sparse representation, memory is not allocated
for every bit independent of the state of that bit. Instead, the non-zero entries are stored only, along with their
coordinates. Therefore, we can save space as long as not most of the data is non-zero. In our case, since
the data is binary, we can go further and always use a compressed encoding, since always either the minority of the 
data will be 1, or it will be 0. We can store an additional bit to represent whether we are storing the locations of 
0s or 1s depending on what best suits.

\subsubsection{Combining Coefficients}

What remains for our quantum emulator to be useful is a way for coefficients to combine when they become of the same sign
due to some operation in the process of executing an algorithm.
We would only combine coefficients that are either both imaginary or both real,
and correspond to the same ket. Let's see how we would combine two coefficients 
with potentially opposite signs.

We would have a state $(-1)^{s_x} x\ket{00} + (-1)^{s_y} y\ket{01} + (-1)^{s_w} w\ket{10} + (-1)^{s_z} z\ket{11}$ maintained
on one oscillator, and $(-1)^{s'_x} x'\ket{00} + (-1)^{s'_y} y'\ket{01} + (-1)^{s'_w} w'\ket{10} + (-1)^{s'_z} z'\ket{11}$ maintained on the other.
In the end we would want $((-1)^{s_x} x + (-1)^{s'_x} x')\ket{00} + ((-1)^{s_y} y + (-1)^{s'_y} y')\ket{01} + ((-1)^{s_w} w + (-1)^{s'_w} w')\ket{10} + ((-1)^{s_z} z + (-1)^{s'_z} z')\ket{11}$.

But what we have represented in the analog module with $a, g$ are the strengths $|x|^2, |y|^2, |w|^2, |z|^2$ and $|x'|^2, |y'|^2, |w'|^2, |z'|^2$. In order to extract them we need 
to follow certain steps. In the next chapter, we will introduce circuits that extract coefficients from the analog module for us. These can be used to find
$x, x', y, y', w, w', z, z'$. Then, we would determine the new strengths by adding or subtracting each of the two complementary coefficients with each other, taking the absolute value, 
and squaring the result. This can be done with a simple arithmetic circuit. Then, with the new strengths, we would run the inverse arithmetic circuit of the relevant measurement circuit we used
to find the coefficients to encode the new strengths into new values of $a, g$.

We will need to combine coefficients after operations like Hadamard, which can turn one ket into two. This is where we cannot get around some of the inefficiency that is typical of emulators 
and simulators. It fits with our observation from preliminary simulations that show the most time is spent when executing operations involving the creation of superpositions.

\subsection{Capabilities and Limitations of the
Scheme}\label{capabilities-and-limitations-of-the-scheme}

As a rough back-of-the-napkin upper bound on the number of qubits we can
represent with the \textasciitilde4580000 states we can achieve with
the AX7MAF1 \cite{abracon_2019}, we can say that it can be no more than
\(Q = \frac{log(4580000)}{log(2)}\overset{\cdot}{=}22\) qubits.
Interestingly, 20 qubits is the number given by La Cour, Ott, Starkey
and Wilson in their paper \cite{la_cour_classical_2016} as the practical amount of qubits possible to
maintain on a single chip as well. Of course, this doesn't take into
account the graininess of the representation or any superposition. It is
just an upper bound. 

However, if we look to higher end oscillators,
there exist CMOS chips that go up to 300 GHz \cite{razavi_300_ghz_2011}. Assuming a high
ppm, perhaps we could represent more states for a price.

The stable precision of the LMH3401 op amp \cite{texas_2014} we have been 
discussing as an example is $\pm 3.2mV$, and so we could only expect to encode 
$\approx 3333333$ values of $a$ using these, which is more limiting than the oscillator.
However, we are not married to the LMH3401 and there exist op amps, for example an ADL5569 \cite{analog_2018},
with a stable precision of $\pm 1.5mV$. In this case the oscillator is again limiting.

\textbf{High Speed Analog Module Cost}

A pair of HSAMs in tandem with a pair of oscillators like the AX7MAF1
will be capable of maintaining $\approx 20$ qubits. As
mentioned, the AX7MAF1 costs roughly \$36.10. High precision LMH3401 op
amps cost \$14. However, more standard op amps may be selected which
have lower slew rates. For example, we might sacrifice encoding speed to
achieve a lower price point by instead using LM358s \cite{industry_standard_2024} which have only a
\(0.6\frac{V}{\mu 2}\) slew rate but cost \$0.16 each.

The entire cost of the analog module's components might be
estimated by:

cost = (oscillators * \$36.10 + op amps * (\$0.16 to \$14) + discrete
components * \$0.01) * 2

= (\$36.10 + 47 * (\$0.16 to \$14) + 170 * \$0.01) * 2

= \$90.64 to \$1391.60.

Low-end cost per qubit is then:

\[\frac{cost}{qubit} = \frac{\$ 90.64}{Q} = \frac{\$ 90.64}{22}\]

High-end cost per qubit is then:

\[\frac{cost}{qubit} = \frac{\$ 1391.60}{Q} = \frac{\$ 1391.60}{22}\]

For example, the low-end cost per qubit is roughly \$4.12. The high-end
cost per qubit is \$63.25.

\subsection{Implementation of Some Clifford Operations}\label{implementation-of-some-clifford-operations}

The implementations of the fundamental Pauli operators will need to take
into account:

• The qubit being acted on

• The topology of the qubit system of which the qubit is a part

However, we will first look at the approach to solving the analog module
implementation of an operator on a two qubit system in order to gain an
intuition for the approach.

\textbf{Quantum Operators as Control Systems}

Given that the analog operator modules are memoryless (excepting their
input voltages perhaps being buffered), a repeatable algorithm for
generating analog circuit implementations directly from quantum operator
matrices can be derived. Like in the Schrödinger Picture \cite[p. 81-82]{sakurai_modern_2014} of a quantum
system, an operator is applied to a state and has the form and function
of a differential operator. The function of a quantum operator U in our
system will be equivalent to the set of the convolution products of U
\RL{'}s decomposed Pauli operators\RL{'} corresponding analog
implementations acting on each qubit. Recall that the convolution
product CP of two functions of time is \cite{PC_221}:

\[CP = \int_{0}^{t}f(t)g(t - \tau)d\tau\]

In the s domain, which is the space that the Laplace transform takes
differential equations, the convolution product is simplified to
straight multiplication.

\[L(f)(s) = F(s) = \int_{o}^{\infty}f(t)e^{- st}dt,s > 0\]

\[L(g)(s) = F(s) = \int_{o}^{\infty}f(g)e^{- sg}dg,s > 0\]

\[CP = f*g\]

A transfer function T(s) is the Laplace transform of the impulse
response of a linear time-invariant system (when the initial conditions
are set to zero). These are practically non-existent in the real world.
However, as stated by Richard Feynman, ``Linear systems are important
because we can solve them'' \cite{feynman_1989}. A linear system satisﬁes homogeneity,
superposition and time invariance requirements. A linear system can also
only consist of six operation types:

\begin{itemize}
  \item multiplication
  \item differentiation
  \item integration
  \item addition
  \item division (multiplication by inverse)
  \item subtraction (addition of negative)
\end{itemize}

These are the operations we have in our arsenal when defining a transfer
function. The Laplace transform of the direct delta signal is 1. We also
know some other handy transforms of transfer functions:

\begin{center}

\begin{tabular}{|c c|}
    \hline
    f(s) & L(f)(s) \\
    \hline\hline
    $\delta$ & 1 \\
    \hline
     $1$ & $\frac{1}{s}$ \\
     \hline
     $t^n$ & $\frac{n!}{s^{n+1}}$ \\
     \hline
     $sin(at)$ & $\frac{a}{s^2+a^2}$ \\
     \hline
     $cos(at)$ & $\frac{s}{s^2+a^2}$ \\
     \hline
     $e^{at}$ & $\frac{1}{s-a}$ \\
     \hline
     $e^{at}sin(bt)$ & $\frac{b}{(s-a)^2+b^2}$ \\
     \hline
     $t^n e^{at}$ & $\frac{n!}{(s-a)^{n-1}}$ \\
     \hline
\end{tabular}

\end{center}

The transfer function T(s) of an analog operator circuit will have at
most two inputs, and two outputs.

Any transfer matrix with this input / output dimensionality is
expressible in a number of forms including the following.

\[\frac{d}{dt}\left\lbrack \begin{array}{r}
x_{1} \\
x_{2}
\end{array} \right\rbrack = \begin{bmatrix}
a_{00} & a_{01} \\
a_{10} & a_{11}
\end{bmatrix}\left\lbrack \begin{array}{r}
x_{1} \\
x_{2}
\end{array} \right\rbrack + \begin{bmatrix}
b_{00} & b_{01} \\
b_{10} & b_{11}
\end{bmatrix}\left\lbrack \begin{array}{r}
u_{1} \\
u_{2}
\end{array} \right\rbrack\]

Where the two outputs would be:

\[\left\lbrack \begin{array}{r}
y_{1} \\
y_{2}
\end{array} \right\rbrack = \lbrack( \neq 0)R\rbrack\left\lbrack \begin{array}{r}
x_{1} \\
x_{2}
\end{array} \right\rbrack\]

Let \(c = \lbrack( \neq 0)R\rbrack\) Then, see that \(c(sI - A)^{- 1}B\)
is equivalent to the transfer function in the s domain.

\[x = Ax + Bu\]

\[y = cx\]

\[(sI - A)x = Bu\]

\[cx = c(sI - A)^{- 1}B = T(s)\]

The matrix format is appealing since it deals directly with the
transformation in a vector space, and the standard fractional transfer
function format of T(s) is appealing since it will be more directly
relatable to a circuit composed of a set of discrete analog components.

The variable s (of the described s domain) is directly representative of
j$\omega$ component, the imaginary unit j times the angle of frequency of the
signal passing through an electrical component \(\omega_{component}\).
With the knowledge that the impedance of a capacitor is
\(Z_{C} = - j\frac{1}{\omega_{C}C}\) and the impedance of an inductor is
\(Z_{L} = j\omega_{L}L\), we can relate a transfer function directly to
an analog circuit composed of discrete components. The are a variety of
control system techniques that may be used to design and optimize a
circuit. However, in our particular case, any quantum operation can be
built by combining a few fundamental operator circuits.

\textbf{$X_{2}$ Gate}

An X gate applied to the second qubit would be equivalent to adding
\(\frac{V_{max}}{4}\) to \(V_{g_{i}}\) when the initial state is within
an odd vertex group (odd quadrant of the \(g\) range), and subtracting
\(\frac{V_{max}}{4}\) from \(V_{g_{i}}\) when the initial state is
within an even vertex group.

We can use a simple voltage divider to attain \(\frac{V_{max}}{4}\).
Operational amplifier, subtractor, adder and comparators can do the
rest. Consider the following circuit which identifies the vertex group
of the initial state, effectively performing a ``vertex group
measurement'' which does not impact the measured state. Let this circuit
be denoted
\(M_{v}\).\begin{center} 
\includegraphics[width=2.60428in,height=2.50322in,alt={VGE\_module.png}]{VGE\_module.png}
\end{center}

The outputs of the circuit have 1 $\omega$ resistors in their places here for
simulation purposes.

\begin{center} 
\includegraphics[width=3.17474in,height=1.91335in,alt={practical\_vertex\_group\_extractor.png}]{practical\_vertex\_group\_extractor.png}
\end{center}

The LT1226 \cite{linear_1992}  was chosen due to its low propagation delay (5.5ns).
Simulating the response of the circuit to a DC voltage sweep using
LTSpice demonstrates its
function.\begin{center} 
\includegraphics[width=3.3602in,height=1.98642in,alt={VGE\_sim.png}]{VGE\_sim.png}
\end{center}

This circuit can be used to implement the X gate applied to the second
qubit using a hybrid digital / analog
circuit.\begin{center} 
\includegraphics[width=6.26389in,height=1.99958in,alt={X\_2.png}]{X\_2.png}
\end{center}

\paragraph{\texorpdfstring{\(X_{1}\)
Gate}{X\_\{1\} Gate}}\label{x_1-gate}

An X gate applied to the first qubit will have a very similar circuit to
the X gate applied to the second qubit. However, there will be an
additional stage since the surface group must also be taken
into account.

\begin{center} 
\includegraphics[angle=90,height=7in,alt={X\_1\_corrected.png}]{X\_1\_corrected.png}
\end{center}

This assumes that the vertex groups are divided into certain divisions.
The first vertex group would be divided into 6 divisions of equal size:

\begin{center} 
\includegraphics[width=5.26389in,alt={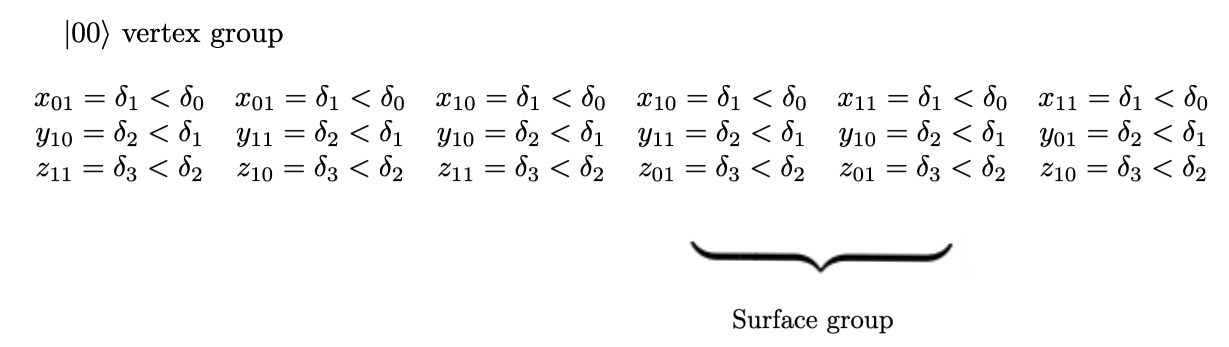}]{divisions.png}
\end{center}

which we can denote by taking the coefficients' subscripts and placing
them in a matrix like:

\[
\begin{bmatrix}
    01 & 01 & 10 & 10 & 11 & 11 \\
    10 & 11 & 10 & 11 & 10 & 01 \\
    11 & 10 & 11 & 01 & 01 & 10
\end{bmatrix}
\]

Then we would have the following.

$\ket{01}$ vertex group

\[
\begin{bmatrix}
    00 & 00 & 11 & 11 & 10 & 10 \\
    11 & 10 & 00 & 10 & 11 & 00 \\
    10 & 11 & 10 & 00 & 00 & 11
\end{bmatrix}
\]

$\ket{10}$ vertex group

\[
\begin{bmatrix}
    11 & 11 & 01 & 01 & 00 & 00 \\
    00 & 01 & 00 & 11 & 11 & 01 \\
    01 & 00 & 11 & 00 & 01 & 11
\end{bmatrix}
\]

$\ket{11}$ vertex group

\[
\begin{bmatrix}
    10 & 10 & 00 & 00 & 01 & 01 \\
    01 & 00 & 01 & 10 & 10 & 00 \\
    00 & 01 & 10 & 01 & 00 & 10
\end{bmatrix}
\]

It should be clear that the analog modules i.e. the operational
amplifiers serve to perform the translation that corresponds to the
manipulation of the coefficients of the kets in the quantum state. What
matters is that the state space is mapped onto divisions of \(g\) in
such a way as to make this operation simple, and that the divisions are
further defined by a space filling function that is repeated identically
in each division with only the variables swapped. The hybrid analog /
digital nature of the circuits is a byproduct of the way in which we are
encoding the basis state bitstrings and their coefficients into one
signal.

\textbf{Controlled NOT Gate}\label{controlled-not-gate}

In the case we are using only one pair of analog modules to maintain a
quantum system, applying a CNOT will be a single voltage shifting
operation similar to a normal NOT (op amps can do it in one shot)
applied on each analog module.

The operation is equivalent to a pure \(X_{2}\) when the vertex group is
either \(|10\rangle\) or \(|11\rangle\). When the vertex group's
bitstring begins with a 0, then we would use similar circuitry to the
\(X_{1},X_{2}\) gates to:

\begin{itemize}
  \item Add Vss/12 if \(g\) is in the first sixth of the vertex group
  \item Subtract Vss/12 if it is in the second sixth
  \item Add Vss/4 if it is in the third sixth
  \item Add Vss/12 if it is in the fourth sixth
  \item Subtract Vss/12 if it is in the fifth sixth
  \item Subtract Vss/4 if it is in the final sixth
\end{itemize}

The circuit itself would be constructed very similarly to the last two,
but is larger and so to conserve space it is not included. Let's call
the circuit that performs the above a ``half CNOT''. Then the full CNOT
would be the following.

\begin{center} 
\includegraphics[width=3.65349in,height=2.41346in,alt={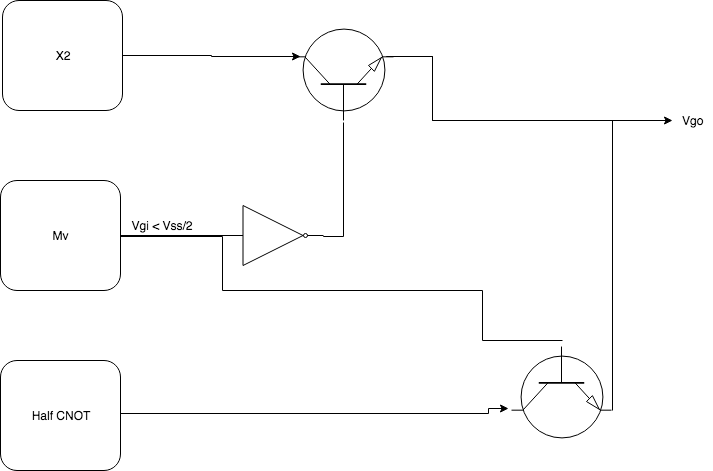}]{FullCNOT.png}
\end{center}

When the initial surface group is 3, then the reflection occurs within
the surface group. Otherwise, it happens between surface groups.

\textbf{Z Gate}

The Z gate applied to both qubits will perform a sign flip to each basis
state coefficient where exactly one of the qubits has a value of 1. It
will negate itself when applied to a basis state with a value of 1 for
both, and have no effect when applied to a state in which both values
are 0. If applied to a valid state, the Z gate will have no effect on
its measurement probability and therefore will consist of a circuit with
an entirely reactive impedance.

In the case of a two qubit system we have 256 possibilities and 256
significant measurable phase shift ranges in the output frequency
oscillation into which the phase and sign of each quantum state
coefficient is encoded. Let the possibilities be ordered according to a
binary count where the variables

are the imaginary and sign flags on each state, ordered by the basis
states\RL{'} quantum bit strings. Then the Z gate will have the effect
of shifting the phase to measurable ranges based on the previous range.

We would want to be able to shift between
\(XX0X0XXX \leftrightarrow XX1X1XXX\) and
\(XX0X1XXX \leftrightarrow XX1X0XXX\). This would require 4 unique
shifts. A unit is exactly \(\frac{2\pi}{256}\) rad. Therefore we need to
design capacitive circuits with overall phase shifts of
\(\frac{3\pi}{16}\) rad, \(- \frac{3\pi}{16}\) rad, \(\frac{5\pi}{16}\)
rad and \(\frac{- 5\pi}{16}\) rad. Let the frequency of the control
signal be fixed. Say we fixed the resistance of the Z gate to 1 $\omega$. Then
a phase shift of \(- \frac{3\pi}{16} = \gamma\) rad could be achieved by
a capacitor of the common value 10µF if the frequency was chosen to be
˙=149.660 KHz. Using 10µF as a reference instead of another common value
such as 1µF was chosen since this gives us the ability to make use of
more of the range of common capacitor and inductor ratings (as you will
see becomes important in a moment, 0.44µF capacitors may be inexpensive
and abundant, but 0.44µH inductors are not).

\[\gamma = - \frac{3\pi}{16} = tan^{- 1}(\frac{- 1}{\omega \cdot 10 \cdot 10^{- 6}})\]

\[\omega = 149.660KHz\]

Since the phase shift of a capacitive circuit is given by:

\[tan^{- 1}(\frac{reactance}{resistance}) = \frac{\frac{- 1}{\omega C}}{R}\]

Which in our case is equal to:

\[= tan^{- 1}(\frac{- 1}{\omega \cdot C})\]

The phase shift of \(- \frac{5\pi}{16}\) rad could then be achieved
using a capacitor of roughly 4.46µF .

\[\gamma = - \frac{5\pi}{16} = tan^{- 1}(\frac{- 1}{149660 \cdot C})\]

\[C = 4.46\mu F\]

While capacitors cause the phase of a signal to lag, inductors have the
reverse effect. So, we can use inductors for the positive phase shifts.

\[\gamma = \frac{3\pi}{16} = tan^{- 1}(149660 \cdot L)\]

\[L = 4.46\mu H\]

\[\gamma = \frac{5\pi}{16} = tan^{- 1}(149660 \cdot L)\]

\[L = 10\mu H\]

To give an idea of the feasibility of this circuit, consider that 8000
10µH inductors with a form factor of 0.6 mm x 0.65 mm x 1.1 mm can be
ordered for \$1.27 apiece form Mouser Electronics. 8000 surface-mount
capacitors of 10µF with form factor 0.8 mm x 0.16 mm x 0.85 mm can be
ordered for \$0.35 apiece from Mouser as well. 4.5µF caps and 4.5µH
inductors are also available in the same price range.

We can achieve the application of the correct phase shift by using a
circuit like the Analog Devices AD8302 \cite{analog_2018}  which accepts two signals and
outputs a voltage proportional to their relative phase shift. We can use
a set of comparators that each control a transistor to select the
correct shifted signal to be sent to the HSAM based on the range of the
phase proportional voltage.

Applying Z to only the first qubit would be achieved by applying a phase
shift of 136 units, a shift of 120 units, a phase shift of -120 units,
or a phase shift of -136 units. Our passive operator components for this
operation would be 2 capacitors and 2 inductors, again. We would see a
similar pattern for the Z gate applied to the second qubit.

Note that the DC voltage drop due to any resistance of the Z module must
be compensated for in its output.

\textbf{Y Gate}

The Y gate can notably be constructed by applying iXZ. We have already
deﬁned the Z and X operations. Therefore, we only need to worry about
the application of i. The application of i to any two quits state with
flag bits FB will be the following, where the upper half of the table
includes circuit components that act on the ﬁrst of the dual
oscillators, and the lower half of the table has elements that operate
on the second. Notice how this provides an opportunity for this hardware
to be reused by both modules rather than duplicated.

\begin{center}
 \begin{tabular}{|c c c|} 
 \hline
 Shift units & Capacitance & Inductance \\ [0.5ex]
 \hline\hline
 85 & 3.79 $\mu$F & 0 \\ 
 \hline
 83 & 3.36 $\mu$F & 0 \\ 
 \hline
 77 & 2.21 $\mu$F & 0 \\ 
 \hline
 75 & 1.85 $\mu$F & 0 \\ 
 \hline
 53 & 0 & 24.1 $\mu$H \\ 
 \hline
 51 & 0 & 20.2 $\mu$H \\ 
 \hline
 45 & 0 & 13.3 $\mu$H \\ 
 \hline
 43 & 0 & 11.8 $\mu$H \\ 
 \hline
 -43 & 3.79 $\mu$F & 0 \\ 
 \hline
 -45 & 3.36 $\mu$F & 0 \\ 
 \hline
 -51 & 2.21 $\mu$F & 0 \\ 
 \hline
 -53 & 1.85 $\mu$F & 0 \\ 
 \hline
 -75 & 0 & 24.1 $\mu$H \\ 
 \hline
 -77 & 0 & 20.2 $\mu$H \\ 
 \hline
 -83 & 0 & 13.3 $\mu$H \\ 
 \hline
 -85 & 0 & 11.8 $\mu$H \\ 
 \hline
 \end{tabular}
\end{center}

\textbf{Measurements}

An advantage of simulating or emulating quantum systems is that system
properties which are normally not measurable can be measured without
disturbing state. Let us first look at some special cases that we know
and understand.

The scheme laid out thus far also begs the question whether we can
efficiently extract any information about the system. Since a complex
encoding of information is assumed by the gates, does this not simply
put the complex decoding off until we want to perform a measurement?

However, we will show that the measurement can also be divided into
cases where the circuits to perform them are often very simple and
powerful, like those of the gates.

\textbf{Chief States}

When a state is on a chief curve of a particular ``chief basis state''
in the probability simplex, we know that the coefficients corresponding
with each other observable state must be the same. We also know that the
coefficient corresponding with the chief observable state is related to
the other coefficients by the following, where \(p \leq 1\) is the
fraction of the probability spectrum maintained in a simplex. See that
\(C_{other}\) and \(C_{chief}\) are related by an elliptic curve.

\[C_{chief}^{2} + 3C_{other}^{2} = p\]

\[C_{other} = \sqrt{\frac{p - C_{chief}^{2}}{3}}\]

Note we omit the $\pm$ sign since we are only concerned with the magnitudes of the coefficients,
and the signs exist in a separate module.

If p = 1, the curve would look like the following segment of an
ellipse:

\begin{center} 
\includegraphics[width=3.06851in,height=2.94122in,alt={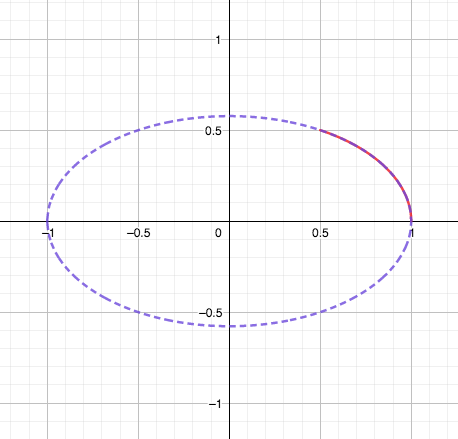}]{elliptic.png}
\end{center}

Recall that two simplexes may be used, one for imaginary and one for
real coefficients\RL{'} components. In the case that only one simplex
has a state with \(a > 0\), p will equal 1 for that simplex and 0 for
the other. In the case that both simplexes maintain coefficients with
significant magnitude, then p will be 1 minus the sum of the squares of
the coefficients maintained in the other simplex. Since \(a\) represents
the distance of a state from the uniform superposition state, we can
determine the magnitude of a coefficient directly from \(a\). In the
case that a state is on a chief curve, the coefficient will be
proportional to the value of \(a\) divided by half the precision in
\(a\).

Then the following circuit would produce a voltage proportional to the
coefficient \(C_{chief}\) of the chief state when the maintained state
is on its chief curve. \(V_{p}\) is a voltage representing the number of
identifiable values of \(a\) due to the precision in \(a\). For example,
if the precision in a were 1\%, Vp would be 100 volts. See that this
would output a value of \(V_{coeff}\) between 0.5 and 1 since
\(V_{coeff}\)
is equivalent to
\(\frac{1}{2}(\frac{V_{a}}{\frac{1}{2}V_{p}}) + \frac{1}{2}\).

\begin{center} 
\includegraphics[width=1.5416in,height=2.21044in,alt={chief\_coeffs\_naive.png}]{chief\_coeffs\_naive.png}
\end{center}

However, it is clear that this circuit is not practical. Addition of
large numbers might be practical to implement since our circuits can use
an adjustable voltage unit \(2V_{\frac{1}{2}}\) when doing addition or
subtraction, but when multiplying or dividing a voltage the multiplier
or divisor need to have actual voltage values corresponding to their
computational value (these volt can\RL{'}t be scaled to use a unit of
\(2V_{\frac{1}{2}}\) . A \(V_{p}\) of 100 volts is the primary concern.
So instead, we must employ a different approach to finding the
coefficients\RL{'} values.

We need the range of the output \(V_{coeff}\) to fill the allowable
domain of the elliptic curve {[}0.5, p{]}. The logistic equation know as
the sigmoid function is ideal for this purpose.

\[\sigma(cx) = \frac{1}{1 + e^{- cx}}\]

The sigmoid evaluated at different values of c yields a family of
curves. c should be selected based on the range of \(V_{coeff}\) and the
precision in \(a\). Say the precision in \(a\) were 100. Then we want
\(V_{coeff}\) to have an approximate value of
\(\frac{1}{2}(\frac{V_{a}}{50}) + \frac{1}{2}\) . A value
\(c = 0.1 \cdot 2V_{\frac{1}{2}}\) fits this purpose since
\(2V_{\frac{1}{2}}\) is our computational voltage unit (CVU). The
Laplace transform of the sigmoid function yields a transfer function
that can be implemented as a PID controller. L denotes the Laplace
transform and \(\mathbb{D}\) denotes the digamma function.

\[L\lbrack\sigma(ct)\rbrack = \frac{\mathbb{D}(\frac{s}{2c}) - \mathbb{D}(\frac{c + 2}{2c})}{2c}\]

So, in our case we will introduce a transfer function \emph{Sigmoid}(s),
where all numbers are given in CVU:

\[Sigmoid(s) = \frac{\mathbb{D}(\frac{s}{0.2}) - \mathbb{D}(11)}{0.2}\]

So \(V_{coeff}\) can simply be obtained
by:\begin{center} 
\includegraphics[width=1.24372in,height=1.65487in,alt={chief\_coeffs.png}]{chief\_coeffs.png}
\end{center}

Then the other coefficients can be calculated using a few components
including a root extractor circuit like that described in Texas
Instruments\RL{'} AN-31 application note \cite{texas_2020}. \(V_{3}\) is 3 volts,
\(V_{pr}\) is the fraction of the probability spectrum maintained in the
simplex.

\begin{center} 
\includegraphics[width=2.8564in,height=3.71888in,alt={c\_other.png}]{c\_other.png}
\end{center}

\textbf{Edge States}

States that are directly on edges of the simplex are easy to solve.
\(a\) is always \(a_{max}\) for edge states. The coefficients for basis
states that are not vertices of the particular edge are zero. The
remaining coefficients \(C_{0}\) and \(C_{1}\) are governed by:

\[C_{0}^{2} + C_{1}^{2} = p\]

\[C_{0} = \sqrt{p - C_{1}^{2}}\]

For these states, we have the relationship of a circular curve between
\(C_{0}\) and \(C_{1}\) with a domain of
\(\lbrack\frac{1}{\sqrt{2}},1\rbrack\).\begin{center} 
\includegraphics[width=4.05887in,height=3.40357in,alt={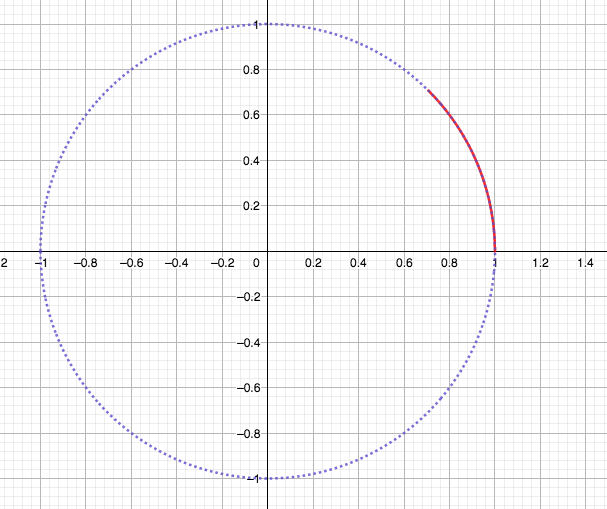}]{circular.png}
\end{center}

In this case we want\(V_{coeff}\) to be between \(\frac{1}{\sqrt{2}}\)
and 1, something like
\(V_{coeff} = (1 - \frac{1}{\sqrt{2}})\frac{V_{g}}{50} + \frac{1}{\sqrt{2}}\).
In a division of \(g\) the edge states would appear periodically and
vary linearly along the edge as we have seen previously. Therefore, we
could choose an appropriate \emph{Sigmoid} function to achieve:

\begin{center} 
\includegraphics[width=3.68383in,height=5.02037in,alt={strongly\_entangled.png}]{strongly\_entangled.png}
\end{center}

\textbf{Edge-Aligned States}

The next most complex case is when the state is on a \(g\) curve which
fills space on a plane that includes two simplex vertices and the
uniform superposition point. In this case we have that two coefficients
are the same, and two are not, like in the edge state case. \(g\) will
have the same bounds as well. However, \(a\) will not be \(a_{max}\),
and the two coefficients that are the same are non-zero.

\[C_{0}^{2} + C_{1}^{2} = p - 2C_{other}^{2}\]

\[C_{other} = \sqrt{\frac{p - C_{0}^{2} - C_{1}^{2}}{2}}\]

\begin{center} 
\includegraphics[width=3.05622in,height=2.38135in,alt={edge\_aligned.png}]{edge\_aligned.png}
\end{center}

\(C_{other}\) is proportional to the value of \(a\) while the ratio of
\(C_{0}\) and \(C_{1}\) is a function of \(g\). If we let
\(V_{coeff}\)represent the coefficient \(C_{other}\) , then we still
have that
\(V_{coeff}\overset{\cdot}{=}\frac{1}{2}(\frac{V_{a}}{50}) + \frac{1}{2}\)
(the circular relationship in the horizontal plane of the graph above).
The remaining coefficients will be related by
\(\lbrack 2(\frac{1}{2}(\frac{V_{a}}{50}) + \frac{1}{2})^{2} - p\rbrack + C_{0}^{2} = - C_{1}^{2}\).
We would use similar elements to the previous circuits to draw this one,
but larger circuits will be omitted to save space.

\textbf{General States}

In the most general case, where there is no short-cut, we need to divide $a, g$ into partitions
according to the divisions described in earlier sections to see where they point to in the tetrahedral 
sphere. The offset into a particular partition would be taken as the domain of the repeated space-filling 
curve, and the final state would be fully determined by that point.

\subsection{Hybrid Digital / Analog Processor Design}

The informational flow diagram of a single, minimalistic processing unit
based on the approaches described thus far might be the following:

\begin{center} 
\includegraphics[width=4.05438in,height=4.94999in,alt={processor\_arch\_updated.png}]{processor\_arch\_updated.png}
\end{center}

The purple and grey blocks represent memoryless hybrid analog and
digital circuits. The purple are pure phase shifts while the grey also
act on DC voltage. The orange represent purely analog circuits. The
green represent the DC component of the oscillator and HSAMs described
in detail in the previous subsection. The blue represents the AC
component of the oscillator and HSAMs.

Note the digital and analog inputs and output to the processing module.
The first input of O bits\RL{'} purpose is to select a grey operation to
be applied, meaning the grey modules must each have a unique bit
address. The number of bits will therefore be
\(O = \frac{operations}{log(2)}\) bits. The input of \(Q\) bits is used
in the creation of quantum states. Each bit directly represents the
state of its corresponding qubit. For example, during the creation of a
state \(|1001011010\rangle\), there would be \(Q = 10\) bits with the
values 1001011010, the state\RL{'}s bitstring. The analog input is also
used during state creation. The outputs simply provide access to the
data in the HSAM / oscillator.

\textbf{High-level Simulator}

The overall functionality can be simulated using Python. The following
code simulates the processor\RL{'}s behaviour, with minimal use of
libraries and pre-built functions. This simulator also provides insight
into the amount of memory and resources that would be required to
perform an algorithm.

Use of the simulator can be summarized into the following use cases. The
simulator is packaged as a library, and the library API is written with
the intent of reflecting the experience of solving a quantum circuit in
Dirac notation. Therefore, what a user of the simulator must provide to
declare a quantum state are the coefficients of its composite basis
states (the square root of each basis state\RL{'}s probability of being
observed) and the bitstring representing the values of each qubit in
that basis
state.

\begin{center} 
\includegraphics[width=4.13668in,height=1.87916in,alt={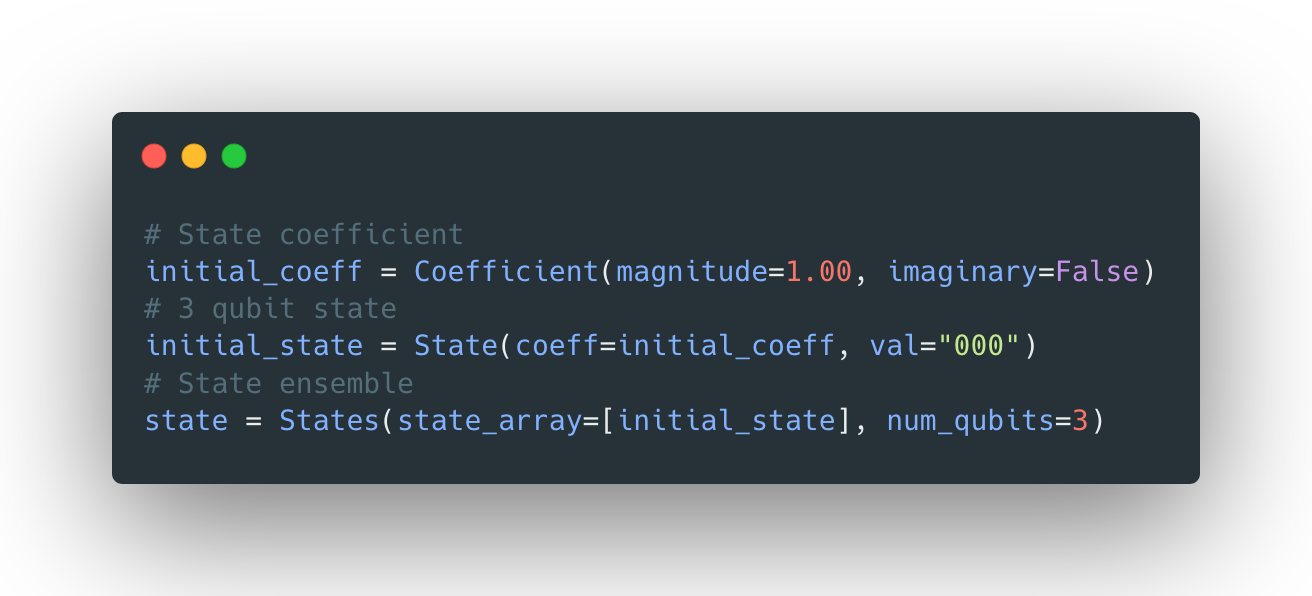}]{carbon9.png}
\end{center}

Supported operations include the Pauli matrices, the Hadamard operator,
and measurement.

\begin{center} 
\includegraphics[width=4.83332in,height=2.46714in,alt={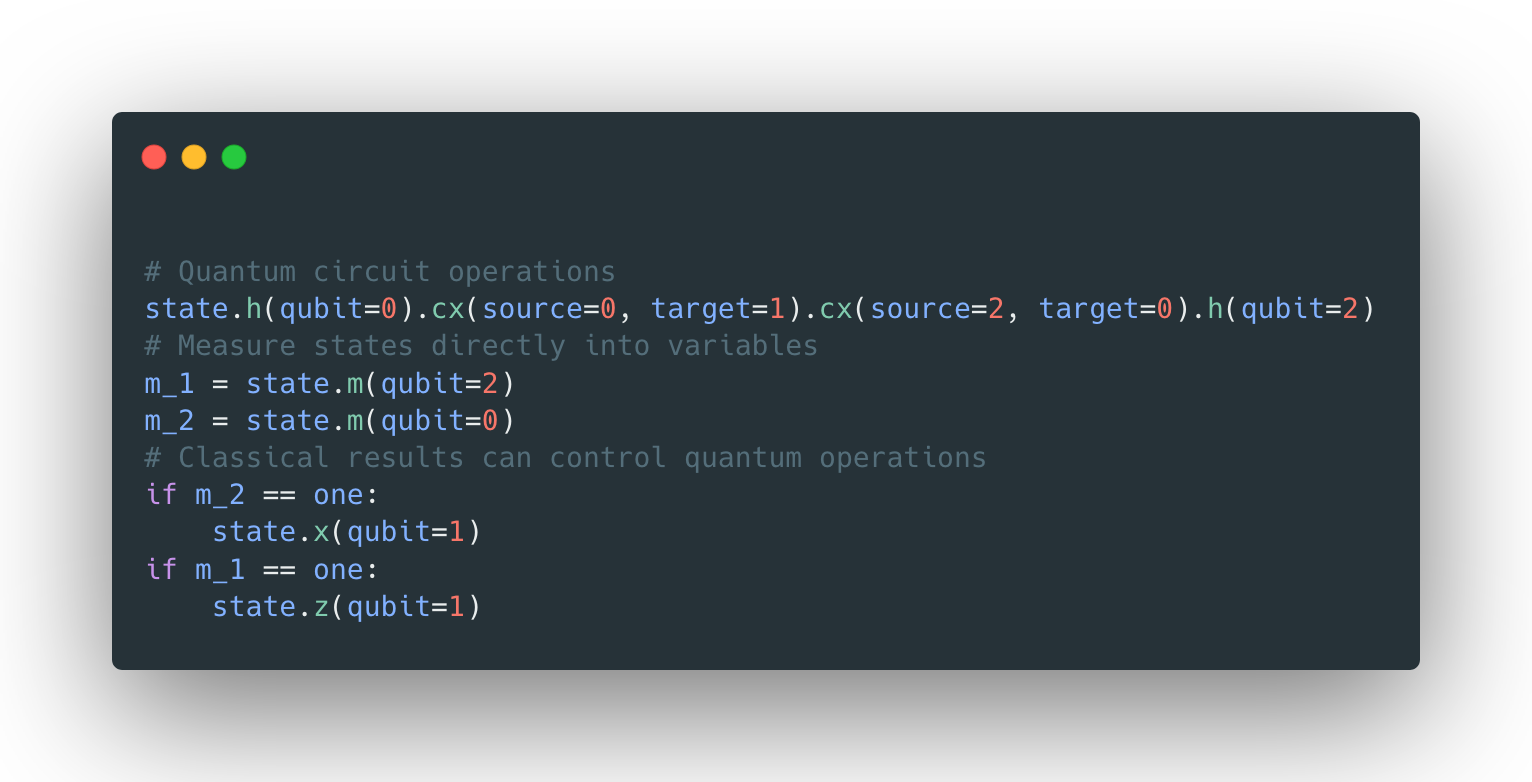}]{carbon10.png}
\end{center}

Insights that can be gained from using the simulator include the
following.

\begin{center} 
\includegraphics[width=4.76237in,height=2.31217in,alt={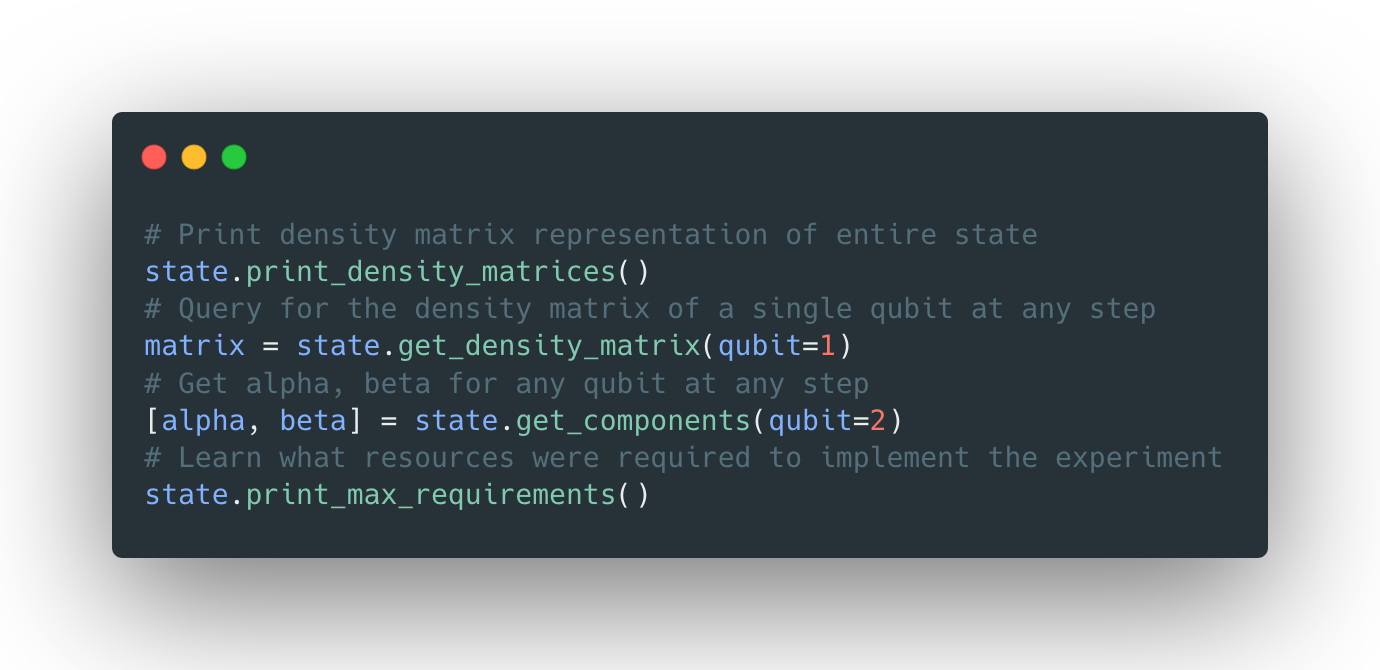}]{carbon11.png}
\end{center}

This simulator has been implemented and tested in order to validate and
optimize the layout of the processor. However, it does not beneﬁt from
the efficiency of the HSAM, or from the inherently faster nature of
bare-metal application specific circuitry.

\textbf{Processor Hardware Requirements}

Having the simulator as a reference makes it a straightforward task to
design the hardware processor, piece by piece. First, we will look at
the first commands given to the simulator in any
simulation:

\begin{center} 
\includegraphics[width=4.55544in,height=1.79547in,alt={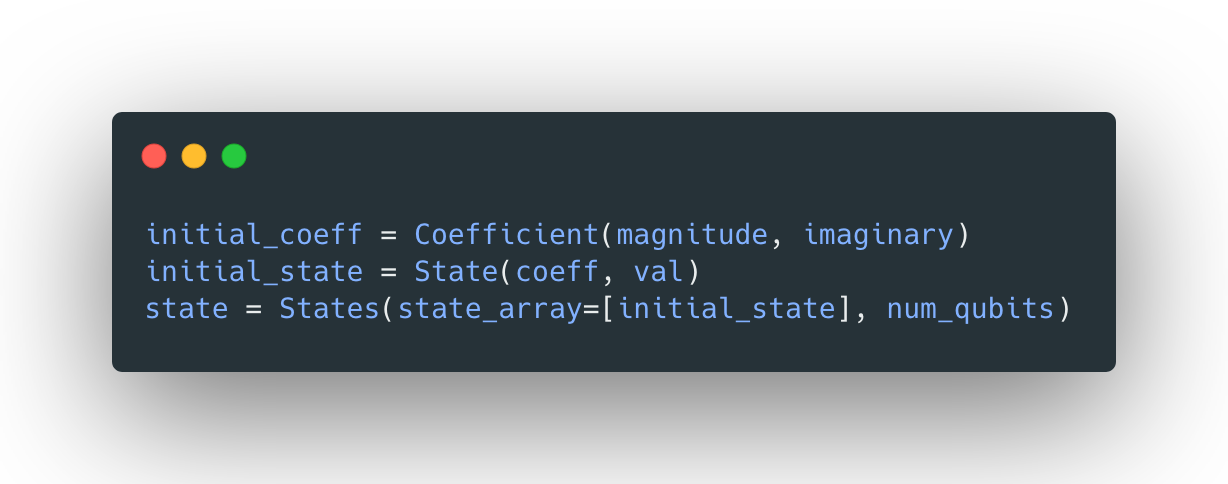}]{carbon12.png}
\end{center}

We know that in the hardware implementation, a coefficient on a basis
state is actualized as the closeness of the emulated state to that basis
state in the probability simplex space. Also, whether the coefficient is
negative or imaginary is maintained as a bit in the flag memory. So, the
``Create Coefficients'' module must be capable of the following tasks:

1. Update \(V_{a}\) and \(V_{g}\) to move to the appropriate distance
from the basis state.

2. Create or update the flag memory for the state.

In order for these to be possible, we must be able to initialize the
analog module and have a system for allocating and managing flag memory
space.

To make the emulator simpler, all basis states will each always be
considered a valid vertex of the qubit state simplex being maintained.
Meaning, we will always emulate all possibilities in the analog module
even if their probabilities of being observed are 0. No shortcuts will
be made during the partitioning of \(g\). When a coefficient creation
begins, the topology of the emulated system will update to match the
number of qubits in the string of \(Q\) bits.

Now consider what it might look like if we did opt to implement a
digital version of the flag processor module.

When the system is initialized, let the number of qubits in the emulated
system directly dictate the size of the flag memory\RL{'}s allocated
space. The allocated space will have a size to account for each of the
negative flags. The allocated space will have a size equal
to \(2^{Q+1}\).
The maximum allocated flag memory size required for one 20 qubit
processor would then be 262144 bytes, or 2097152 bits. The number of
qubits must be stored so that the system is capable of distinguishably
delimiting the allocated space. 5 additional bits will be required to
store this value. Let the allocated flag space always begin in memory
location 0, and the number of qubits be stored in the final 5 bits of
memory. In order to manage this digital memory, we will need to
implement a digital information pipeline.

Operations on the flag bits of n qubits without dependencies on current
qubit states are very simple. For example, a NOT gate does not need to
know what the state of its target qubit is in order to flip its bit. In
terms of digital information, all it needs to do is interchange the flag
bits of each target qubit state with its complement\RL{'}s. This can
certainly be done by a combinational circuit in one clock cycle. The
same is true of a sign flip operation, which simply needs
to call upon the ``negate'' module to mutate the
correct memory locations. Controlled operations are also simple and
should just be applied to every state that has the value 1 in the place
of the controlling qubit.

So, the details of the operations that will affect their timing
performance lie within the implementations of their modules (which are
hopefully achievable as combinational circuits) and in the memory
management mechanism.

\textbf{Digital Memory}

Static SRAM will be chosen for the flag memory since it is generally
more affordable than NVRAM. The downside is that memory values will not
be maintained if the device is powered down, which is acceptable. For
example, we might chose a memory chip like the IS62WV12816EALL \cite{integrated_2013}, a 128K x
16 low voltage, ultra low power CMOS SRAM chip that costs roughly \$1.28
each. The IS62WV12816EALL has a write cycle and access time of 45ns \cite{integrated_2013},
well underneath the time required for the analog module to finish its
part of an operation (10 µs).

Of course, the fastest type of memory is a simple latch. Texas
Instruments\RL{'} SN74HSTL16918 \cite{texas_1997} is a latch array with a 1.9 ns delay for
writes, and no delay for reads due to the nature of the D-latches used.
However, the SN74HSTL16918 \cite{texas_1997} has only 18 bits and costs roughly \$2.75, an
affordable price for latch based memory.

We might instead choose a chip with a parallel access memory interface,
like the one that the IS62WV12816EALL also uses. This is a fast
interface because it is asynchronous. We might chose a chip like the
AS6C6264 \cite{alliance_2009}, a \$2.43 8K x 8 low power CMOS SRAM chip with 64Kb of memory
and write cycle and access times of 55ns. This far surpass the several
micro second write and access times of I2C and SPI interfaced memory.

If we are not worried about having excess memory, we might chose to
splurge and use a single 4Mb memory chip like the IS61WV25616EDBLL \cite{integrated_2020}, a
256K x 16 high speed asynchronous CMOS SRAM chip that costs only \$2.11.
Its write and access delays are only 10ns. This gets us more than the
memory we need, but also costs less than the more precise alternatives.

In any case, we will choose to use parallel asynchronous memory. Let us
assume we went with a single chip similar to the IS61WV25616EDBLL and
have memory to spare. The IS61WV25616EDBLL datasheet \cite{integrated_2020} provides the block
diagram of the chip\RL{'}s functionality, pinout diagram and I/O truth
table.

Since we have an excess of memory, and are not concerned too much with
using space, we can forego memory location optimization and simply
allocate 2 consecutive bits per state. Then the address for a
state\RL{'}s flag bits is simply the state\RL{'}s bit string times 2.
The circuit that calculates a state\RL{'}s flag memory

location is simply a left bit shift of 1 digit on the state\RL{'}s bit
string. However, memory chips like the IS61WV25616EDBLL address words of
16 bits. So, the nth bit will occur in the word at index
\(\lfloor n/16\rfloor\). To calculate the word address, we could shift
the bit address right by 4 digits. See that the direct calculation of a
word address from the state string would be a right bit shift of 3
digits on the state bit string. The location of the bit address within
the word address will then be given by the remainder of the n/16
division, and be given by the bits shifted out of the state bit string
during the right bit shift. Several microprocessors include support for
a command that shifts a bit string through a register in order to retain
the bits pushed off of the end, including the Motorola 68000, an old an
inexpensive processor. However, the 68000 and similar processors do a
lot more than is required for the quantum emulator being designed. So,
we will take a page from their book but not go so far as to use a third
party microprocessor product. We can build something even less costly.

\textbf{Digital Microprocessing Layer Elementary Operation Pipelining
Infrastructure}

The digital pipeline of our processor will be unique. We will not need
to discuss modern processor pipeline architecture since our processor
will be far simpler than even the classic RISK architecture used by
early processors like the M68000 and MIPS processors. Classic pipelines
included only a few stages. The names of the pipeline steps in the MIPS
architecture were ``Instruction fetch'', ``Instruction decode'',
``Execute'', ``Memory access'' and ``Writeback''.

It will be necessary to include a stage for fetching instructions, since
it will be necessary for a user to select each operation to apply. The
fetch step will occur periodically.

The decode step will consist of identifying the operation to be
executed. and determining whether it is safe to trigger the
operation\RL{'}s digital and/or analog circuits. We have a unique
problem to solve in this design since we are executing a set of
relatively lengthy analog operations in which the order of
operations is significant, alongside relatively fast digital operations in
which order is insignificant. Due to this fact, it is ideal to separate
the execution step into digital and analog execution steps. The decode step will
be responsible for taking in an instruction and determining the
underlying digital and analog operations, each of which will be scheduled
separately.

Before we begin the design of the task scheduling mechanism, we will
first design a mechanism for combining operations that takes advantage
of the memoryless nature of the operation modules. Assuming each
operation module\RL{'}s analog circuitry is a signal processor, and each
operation module\RL{'}s digital circuitry is combinational, we might be
able to perform several ''sequential'' operations at once by piping
together the modules\RL{'} circuits in the correct order. for example,
if we wanted to apply a NOT gate to a qubit, and then a Z gate to the
same qubit, the analog NOT circuitry and Z circuitry could be applied
to the \(V_{a}\) and \(V_{g}\) at once. This would be the same as the
convolution of their transfer
functions. We would need to be able to control both the path of Va and
Vg into the first circuit, and the path of the first circuit\RL{'}s
output into the second circuit. This could be simply achieved by
using transistors or relays. Each operation module\RL{'}s analog
circuitry could have an output that is multiplexed to the HSAM\RL{'}s
inputs as well as each of the other analog modules\RL{'} inputs. Then it
is the responsibility of the decode circuitry to compose compatible
steps in the correct order. This achieves a sort of ''sequential''
instruction level parallelism that maintains the order of operations.
The limitation of this scheme would be that each module can only be used
once in the convolution of operations.

Both instruction level parallelism and out-of-order execution can be
taken advantage of in the digital circuitry surrounding the operation
modules. Since the only mutations done by the one pipeline are sign and
phase flips, order does not matter. In fact, any set of sign flips on a
state can be summarized into one flip if the number of flipping
operations is odd, and any even number of sign
flips can be ignored entirely. The behaviour of each of the phase and
sign flip modules can be summarized as
follows.

\begin{center}
 \begin{tabular}{|c c c c|} 
 \hline
 Imaginary input $I_i$ & Sign input $I_s$ & Imaginary output $O_i$ & Sign output $O_s$ \\ [0.5ex]
 \hline\hline
 0 & 0 & 1 & 0  \\ 
 \hline
 0 & 1 & 1 & 1  \\ 
 \hline
 1 & 0 & 0 & 1  \\ 
 \hline
 1 & 1 & 0 & 0  \\ 
 \hline
\end{tabular}

\vspace{5mm}

Single multiplication by i
\end{center}

\vspace{5mm}

\vspace{5mm}

\begin{center}
 \begin{tabular}{|c c c c|} 
 \hline
 Imaginary input $I_i$ & Sign input $I_s$ & Imaginary output $O_i$ & Sign output $O_s$ \\ [0.5ex]
 \hline\hline
 0 & 0 & 0 & 1  \\ 
 \hline
 0 & 1 & 0 & 0  \\ 
 \hline
 1 & 0 & 1 & 1  \\ 
 \hline
 1 & 1 & 1 & 0  \\ 
 \hline
\end{tabular}

\vspace{5mm}

Single negation
\end{center}

In other words, the negation is just a bit flip of the sign bit and can
be used to represent any odd number of negating operations. The
multiplication by i is simply a bit flip of the imaginary flag (potentially achieved indirectly by swapping amplitudes), and also
a negation of the state when the imaginary input is true.

The number of negations actually performed might be summarized in terms
of the number of multiplications by i \(N_{i}\) and negating operations
\(N_{- 1}\) as:

\[(\lfloor\frac{N_{i} + I_{i}}{2}\rfloor + N_{- 1})\text{mod}2\]

The number of imaginary flag flips actually performed might be
summarized in terms of the number of multiplications by i \(N_{i}\) as:

\[\lfloor\frac{N_{i} + I_{i}}{2}\rfloor\text{mod}2\]

We should also consider that any of H, X, Y and Z have no effect when
applied an even number of times in succession to a particular set of
states. In particular, the number of successive gates in this set to
actually apply would be given by the number of similar gates in a row
\(N_{s}\) :

\[N_{S}\text{mod}2\]

Since it is apparent that summarizing and composing computational steps
will be useful, and we are treating the digital and analog elements of operations
as separate pipelines, we will adopt a VLIW architecture. VLIW stands
for ''Very Long Instruction Word'', and indicates a processing
architecture that accepts multiple instructions per instruction word and
schedules each instruction to be processed by one of a number of
parallel pipelines in such a way that there are no conflicts between
instructions.

The initial design includes five operation modules that might be piped
together. So, we will start with 5P5 + 5P4 + 5P3 + 5P2 = 320 unique
convolutions which are permutations of the modules without repetition.
We will use $5^2 = 25$ bits to select a particular convolution, one for
each transistor or relay required. The bits used for specifying a
convolution can be derived from the operation identifier bits in the
instruction word if we assume that operations are listed from left to
right in the order that they are meant to be applied. We will also need
to include bits that identify the target qubit for each operation, and
the source qubit for controlling operations. We will assume that each
operation identifier is followed by a 5 bit qubit identifier, and that a
controlling operation is followed by two 5 bit qubit identifiers; for
the source and target qubits respectively. Next we will solve the
circuit that calculates the permutation of operations to be convoluted
based on the operation bits in the instruction word, assuming that all
the operations target one same set of states.

\vspace{5mm}

\begin{center}
 \begin{tabular}{|c|c|} 
 \hline
 address & operation\\ [0.5ex]
 \hline\hline
 00001 & X \\
 \hline
 00010 & Y \\
 \hline
 00100 & Z \\
 \hline
 01000 & M \\
 \hline
 10000 & H \\
 \hline
 \end{tabular}
 
\end{center}

We can minimize the standard SOP boolean expressions that follow from
this table using the Quine--McCluskey algorithm \cite{floyd_digital_2009}. The following code
implements this optimization algorithm for our particular circuit.

\hspace{-12mm}
\includegraphics[height=7.75in,alt={fig\_5\_7.png}]{fig\_5\_7.png}

\hspace{-15mm}
\includegraphics[height=7.75in,alt={fig\_5\_8.png}]{fig\_5\_8.png}

\hspace{-12mm}
\includegraphics[height=5.5in,alt={fig\_5\_9.png}]{fig\_5\_9.png}



An additional stage of processing to be invoked before this will apply the
optimization of summarizing any H and/or Pauli operations being applied
multiple times in succession. Each adjacent (extended) nibble in the operation bits
of the instruction word will be compared using a simple 5-bit comparator
circuit. If any adjacent nibbles match, then they will be removed by
shifting the contents of the operation bits to the right of the matching
operations into the registers of the adjacent nibbles. This is very easy
to achieve using a simple digital processor with a bit shift function.
It is also implementable using strictly combinational logic.
Let the output of such a combinational circuit be the bit string C. The
optimization of the combinational version was found using the following
code.

\hspace{-5mm}
\includegraphics[height=6in,alt={fig\_5\_10.png}]{fig\_5\_10.png}

\hspace{-5mm}
\includegraphics[height=6in,alt={fig\_5\_11.png}]{fig\_5\_11.png}

\hspace{-12mm}
\includegraphics[height=7.75in,alt={fig\_5\_12.png}]{fig\_5\_12.png}

Notice that these combinational circuits\RL{'} combined function is to take a
set of operations meant to be applied to one set of states and decode
the instruction containing their bitcode identifiers such that they are
optimally executed. However, we have not yet addressed how the decode
step of the pipeline will know which set of basis states a series of
operations is meant to be applied to. So, we must use the qubit
identifiers in the instruction word, the ``instruction target
qubits\RL{'} bits (T)'', to identify the overlaps of targeted states.

It is also true that the analog modules for each quantum gate\RL{'}s
application to each qubit will either be different or take a parameter
identifying the target qubit(s). This is true since gates will have
different effects on the overall state depending on the target(s) of
their application. For example, the X gate will flip the mixed state
within the maintained probability simplex over a different plane or
hyperplane that passes through the uniform superposition state for each
different set of target qubits. The case that the modules take
parameters is vastly more ideal than the other option, so we will assume
that we can set a voltage in order to target a qubit at the analog gate
level. In that case, this parameter voltage must be set before applying
a set of gates to a qubit that may occur together. It may be connected
in parallel to each analog module\RL{'}s circuits. for gates that cannot
take such a parameter, we will assume that the operation is considered a
different operation altogether with a different module address, in which
case it will not cause conflicts within this decode logic and will be
addressed later.

So, we will endeavour to group consecutive operations by qubit in order
to optimize their execution. Let a full instruction word contain no more
than one set of 5 operation and target qubit codes. Let it be assumed
that an instruction word will be present in a set of registers close to
the decode logic at the beginning of the decode step. Then a task of the
decoder prior to the combinational logic we have deﬁned will be to parse
the groups of adjacent operations in the word with common qubits into
groups of at most (and ideally) 5.

These groups will then each be passed to a separate instance of the
combinational decoder logic that
will exist for each of the 20 qubits. We might have duplicates of each
operation that can receive different target qubit parameter voltages at
once and still be piped together for each possible qubit. That would
leave us with 5 × 20 = 100 total analog modules including X, Y , Z, H
and M . The sets of 5 analog modules should each be constantly piped as
every set should always be used as long as there are enough operations
in the long instruction word. When no qubit is selected, the set should
simply be bypassed using a single threshold triggered transistor or
relay.

This suggests that our full instruction word should contain 20 sets of 5
operations, meaning it has 4 × 5 × 20 operation bits (W) and 5 × 5 × 20
target qubit identifier bits (T), meaning it is 900 bits long.

We must also be prepared to consider the addition of more hybrid analog
/ digital modules to the system in the future. The architecture\RL{'}s
capacity for additional sets of modules must not delay the decode
instruction if these extra sets are not installed, and it must be
possible to extend the modules of the system programmatically. This
extension must not be deﬁned in hardware with any hard limitations. For
now, it will suffice to say that the output of the final piped module
set can be programmatically multiplexed to either output directly to the
HSAM, or to output to an analog output pin. The input voltage to the
HSAM must also be exposed as an input pin. We might want to design our
memory management system and also design a machine language before this
problem can be addressed fully.

A register architecture is beginning to be implied by the design. We now
have a 900-bit instruction word and 20 45-bit inputs to separate parts
of the circuitry. It is also evident we will need the machine to be
capable of receiving consecutive instruction words with a certain
frequency and handling them sequentially. For this purpose, it will be
necessary to design a a hardware (or software) digital microprocessor
around our hybrid analog/digital operation pipeline architecture that
implements minimal but ideal control logic. This is a typical exercise
in optimizing control logic that could be performed by any competent
embedded systems designer. The fundamental advantage for quantum
emulation of our design is entirely encapsulated by the work presented
thus far.

The circuitry deﬁned thus far plays the role of the execution modules in
a digital pipeline. A processor built around our execution module might
have no ALU or FPU. Rather, it could have the described operation
pipelining infrastructure and resulting combinational digital logic and
convolutional analog differential operator circuits in their place. This
would allow it to achieve efficient quantum information emulation.
Alternatively, our execution module could be added to supplement the FPU
or ALU in a pre-existing computing system.

\subsection{Optimality Comparison}\label{optimality-comparison}

The optimality of the proposed emulator will be compared to a
prototypical hardware general quantum computing emulator that makes use
of digital FPGA technology. Dlugopolski and Pilch published an
FPGA-based real quantum computer emulator \cite{pilch_fpga_based_2019} design in the Journal of
Computational Electronics.

Dlugopolski and Pilch implemented their own 10 bit floating point
arithmetic in VHDL. They created a quantum computing emulator with the
following architecture.

The quantum state was maintained as a set of 10 bit floating point
numbers, one float for each of the real and imaginary components of each
coefficient. Quantum transformations were implemented by parallel
floating point multiplications acting on input gate registers and inputs
state registers.

Qubit measurement in their design was implemented as Von Neumann
measurement, requiring the following steps.

\begin{itemize}
\item Probability of measuring 0 is computed based on the entire state

\item A pseudo-random real number is generated

\item If the number from step 2. is greater than the probability from step 1., qubit\RL{'}s measured value is set to 1. Otherwise, the qubit\RL{'}s measured value is set to 0

\item Amplitudes of all impossible states (ones where selected qubit\RL{'}s value is different than measured) are set to 0 entire state

  \item All amplitudes are normalized so that \(\sum_{i}(\text{amplitude}_{i})^{2} = 1\)
\end{itemize}

This procedure clearly scales exponentially with the number of qubits
being maintained. However, their implementation is valuable in the sense
that it offloads the complexity of emulating quantum information systems
completely to the amount of digital hardware being thrown at the
problem, and this hardware is being used optimally.

Their study results in the following relationship between hardware ALMs
and representable qubits. To recognize the significance of an ALM,
consider that some Intel Cyclone V FPGAs have 32,000 onboard ALMs \cite{cheung_2019}.

Using Cyclone FPGAs, using Dlugopolski and Pilch\RL{'}s approach would
cost around \$900 to emulate 4 qubits, or \$5000 to emulate 5 qubits.

While the computational speed of their approach is not analyzed in their
paper, it is easy to ﬁnd the FLOPs that commercial FPGAs are capable of
performing. For example, an entry level FPGA might be capable of 4
GFLOPs \cite{bosch_exploiting_2017}.

Since our approach does not explicitly use floating point operations to
emulate quantum computing, we must introduce a comparable performance
metric. The precision and slew rates of the op amps used in our design
will be the limiting characteristics of our approach. Since the op amps
are used only in common configurations that don\RL{'}t depend on their
specific implementations, we have the ability to choose the family of
amps to employ based on these characteristics, and weigh efficiency
against cost.

High precision LMH3401 \cite{texas_2014} op amps have a slew rate of 18000 V / µs and cost
\$14 apiece. This translates to a max propagation delay of roughly 0.5
ns considering that we may have a voltage range of 10 volts.

Taking a look at the critical path through the X1 operation\RL{'}s
circuitry, we can estimate the propagation delay of the entire operation
to be (5 op amps) times 0.5 ns. The power of this is that the amount of
equivalent floating point ops scales with the system size. One X
operation updates the encoded information such that all of the new
coefficients are represented properly in the analog modules, without
incurring any extra propagation delay. This is because the complexity is
in the encoding, which we address by our clever decoding using the
measurement circuits discussed. We only need to be able to choose them
according to some knowledge about the circuit.

Iff we assume the familiar case where we are updating 4 coefficients
with our module, \(5 \times 0.5\) ns would achieve 4 FLOPs, then we
achieve something equivalent to 1.6 GFLOPs. The break-even (rounding up) is the next
system size up, where we achieve 3.2 GFLOPs.

If we notice that the HSAM and oscillator are not actually required
elements of the system for computation, but only required for
information encoding for output, we can estimate the cost of a
computation-only system based on our approach by summing the costs of
the op amps in the quantum operator modules: \$14.55 × (50 op amps) =
\$727.5. This is notably less than the estimated \$900 cost to emulate 4
qubits using Cyclone V FPGAs.

The selected op amp has a 30 mV stable precision. However, by
introducing another, cheaper LMV771 \cite{texas_2010} op amp we can perform common mode
voltage correction for \$0.55 per op amp and achieve a 3 mV precision.

The number of a values encodable can be estimated by 10v
\(a_{max}\overset{\cdot}{=}3,333,333.3\) distinguishable states. Since
this is less than the number of states that are distinguishable in the
oscillator output, this shows that the op amp choice is the limiting
factor for the design\RL{'}s capabilities if the LMH3401 is used.

If we wanted to increase the performance of the op amps, we could choose
a more performant amp, say the ADL5569 \cite{analog_2018}. This op amp has a slew rate of
24000 V / µs and therefore would allow our device to perform with a much
greater efficiency. However, each is around \$28, so the cost of a
device would almost double, increasing to \$1400.

\textbf{Conclusions}

We outlined limitations of naive implementations of analog quantum
computer emulators. For example, the exponential scaling of the state
space with the number of qubits can manifest itself as an exponential
requirement for bandwidth. We partially overcame this issue by encoding
our state partially in an RMS voltage intermediately instead of a
frequency, using the frequency domain at this stage only for encoding
the phase component of the state, which is not relevant or used in the
measurement of the quantum state in the computational basis. The
frequencies used to represent the quantum state in the end are derived
via a mathematical framework inspired by the quantum harmonic oscillator
to ensure that the frequencies exist in a hierarchy that is in a sense
optimal. This is done in a very similar way to how one may simulate n
qubits with a qudit by mapping the basis states onto its energy levels.
In order to take full advantage of this model, a hierarchy of tunable
mathematical constructs analogous to qudits that were guaranteed to have
unique levels from one another were used instead of just the one, to
pack more unique state representations into the bandwidth available.
Visualizations and corresponding circuits for working with this
representation were provided for the case of two qubits. While the
simplicial geometries depicted generalize straightforwardly to n qubits,
they are hard to visualize beyond two qubits as those geometries are
beyond 3 dimensional.

We showed that it is comparatively inexpensive to outmatch the computing
capabilities of existing FPGA based emulators \cite{pilch_fpga_based_2019}, as our approach
offloads the complexity of atomic operations onto the complexity of
decoding that state. However, fast modules were also detailed for
decoding as long as some information is known about the nature of the
state.

Scaling beyond tens of qubits is still impractical due to constraints of
analog hardware module
precision.

\subsection{Paper author
contributions}\label{paper-author-contributions}

Marcus Edwards designed and documented the emulator as well as performed
the related proof of concept experiments before finally writing the
paper.

\subsection{Acknowledgments}\label{acknowledgments}

Thanks to Dr. Shohini Ghose of Wilfrid Laurier University for helpful
discussions on early versions of this work.

\nocite{mermin_quantum_2007, industry_standard_2024, edwards_developing_2019, CP_216, CS_450, pinedo_scheduling_2022, maronese_quantum_2022, floyd_digital_2009, bosch_exploiting_2017, nielsen_quantum_2010, ates_sigmoid_2015, park_sigmoidal_2005, raghunandan_encoding_2006, demrow_settling_1970, larose_overview_2019, Luk2018, baugh_twos_1973, butkovskiy_control_1990, wu_control_2012, ieee_VHDL, feynman_simulating_1982}

\bibliographystyle{plain}
\bibliography{main}

\end{document}